\documentclass[preprintnumbers,article,amsmath,amssymb,floatfix,10pt,prd,onecolumn,
superscriptaddress,nofootinbib]{revtex4}
\usepackage{bbm}
\usepackage{amsfonts}
\usepackage{mathrsfs}
\usepackage{latexsym}
\usepackage[cp1251]{inputenc}
\usepackage{epsfig}
\usepackage{epstopdf}
\usepackage{graphicx}
\usepackage{amssymb}
\usepackage{amsmath}
\usepackage{dcolumn}
\usepackage{bm}
\usepackage{color}
\usepackage{comment}
\usepackage{xcolor}
\begin{document}

\title{\bf Traversable wormholes in the extended teleparallel theory of gravity
with matter coupling}

\author{G. Mustafa}
\email{gmustafa3828@gmail.com}\affiliation{Department of Mathematics, Shanghai University,
Shanghai, 200444, Shanghai, People's Republic of China}

\author{Mushtaq Ahmad}
\email{mushtaq.sial@nu.edu.pk}\affiliation{National University of Computer and
Emerging Sciences, Islamabad,\\ Chiniot-Faisalabad Campus, Pakistan.}

\author{Ali {\"O}vg{\"u}n}
\email{ali.ovgun@emu.edu.tr}\homepage{https://aovgun.weebly.com}

\affiliation{Physics Department, Eastern Mediterranean
University, Famagusta, North Cyprus via Mersin 10, Turkey.}

\author{M. Farasat Shamir}
\email{farasat.shamir@nu.edu.pk; farasat.shamir@gmail.com}\affiliation{National University of Computer and
Emerging Sciences,\\ Lahore Campus, Pakistan.}

\author{Ibrar Hussain}
\email{ibrar.hussain@seecs.nust.edu.pk}\affiliation{School of Electrical Engineering and Computer Science,
National University of Sciences and Technology, H-12, Islamabad, Pakistan}

\begin{abstract}
This study explores the Gaussian and the Lorentzian distributed spherically symmetric wormhole solutions in the $f(\tau, T)$ gravity. The basic idea of the Gaussian and Lorentzian noncommutative geometries emerges as the physically acceptable and substantial notion in quantum physics. This idea of the noncommutative geometries with both the Gaussian and Lorentzian distributions becomes more striking when wormhole geometries in the modified theories of gravity are discussed. Here we consider a linear model within $f(\tau,T)$ gravity to investigate traversable wormholes. In particular, we discuss the possible cases for the wormhole geometries using the Gaussian and the Lorentzian noncommutative distributions to obtain the exact shape function for them. By incorporating the particular values of the unknown parameters involved, we discuss different properties of the new wormhole geometries explored here. It is noted that the involved matter violates the weak energy condition for both the cases of the noncommutative geometries, whereas there is a possibility for a physically viable wormhole solution. By analyzing the equilibrium condition, it is found that the acquired solutions are stable. Furthermore, we provide the embedded diagrams for wormhole structures under Gaussian and Lorentzian noncommutative frameworks. Moreover, we present the critical analysis on an anisotropic pressure under the Gaussian and the Lorentzian distributions. \\\\\\

\end{abstract}

\maketitle

\date{\today}

\section{Introduction}

The crux of today's cosmological paradigm is the wondering of the accelerated expanding universe and its potential causes, as affirmed by various astrophysical tests, which have become a focal point of interest for the recent studies \cite{1,2,3}. In this regard, the principal endeavor was made by Einstein, presenting a notable $\Lambda$CDM model; yet notwithstanding its all excellence and achievement, this model cannot be accepted as a final one \cite{4}. Later on, various propositions have been introduced by the scientists that can be gathered into two sorts of modified propositions: modified matter propositions and the modified curvature propositions. For instance, the tachyon model, the Chaplygin gas, quintessence, and its various variants, quintom, phantom, and so on, are gotten by presenting some additional terms in the matter content and thus are associates of the modified matter proposition cluster \cite{5,6,7,8,9,10,11,12,13}. The other thought is to offer the modification of action of Einstein's General Relativity (GR) by including some additional degrees of freedom. One of the essential adjustments was the Lagrangian density due to the Einstein-Hilbert action incorporating the $f(R)$ function, with $R$ being the Ricci scalar. This theory has been extensively used in the work done by Nojiri and Oditsov \cite{16}, to study dark energy (DE) and its subsequent expedient cosmic expansion. Additionally, the $f(R)$ theory of gravity provides an integrated depiction of the initial phases of the inflationary universe just as the late phases of the expanding universe \cite{Cimdiker:2020enx}. Classical works comprising the modified theory of $f(R)$ gravity by Capozziello et al. can be found in the literature  \cite{CZ1,CZ2,CZ3,CZ4,CZ5,CZ6,CZ7,CZ8}. In a recent work physically viable wormhole solutions in the $f(R)$ theory of gravity with exponential model have been discussed by Mustafa et. al \cite{Uni1}. Non-extremal spherically symmetric instanton wormhole solution in string theory have been discussed by Bergshoeff et al. \cite{PP1}. Maldacena et al. have presented a study of some traversable wormhole geometries in the background of $AdS_2$ gravity \cite{PP2}. Some other notable models include Brans-Dicke gravity, $f(\tau)$ gravity, with $\tau$ being a torsion, generalized Gauss-Bonnet theory of gravity with its further general modifications as the $f(G)$ gravity, $f (R, G)$ gravity, and $f(\tau, \tau_{G})$theory, etc. \cite{17,18,19,20,21,22,23,24,25,26}. One more critical modification of the Einstein gravity to be specific $f(R, T)$ gravity and was anticipated by Harko et al. \cite{27} right around five years prior. The structure of the bivariate function $f(R, T)$, incorporates the connection of the Ricci scalar $R$ and $T$, the trace of the energy-momentum tensor.

Nevertheless, all recently referenced fruitful modified gravity theories which depend on a rudimentary element of metric tensor $g_{uv}$, another
proposition option alternative to Einstein gravity exists, which is recognized as the teleparallel gravity. Proposed by Einstein himself, this theory is equivalent to GR, where the tetrad space is exploited as essential physical variable \cite{8B}. In this continuance, another intriguing modified gravity is acquired by supplanting the torsion scalar $\tau$ with a subjective generic function, known as $f(\tau)$ gravity \cite{9B}. This theoretical approach is considered as quite possibly the most fascinating option as a substitute to GR in light of the straightforwardness of its subsequent field condition. Ahmed et al. \cite{10B}, examined the accumulation cycle of a spherically symmetric black hole under teleparallel $f(\tau)$ gravity. Boehmer and his collaborators \cite{bbb}, explored the possibility that static and spherically symmetric traversable wormhole (WH) geometries are supported by modified teleparallel gravity. Bahamonde et al. \cite{11B}, studied the Lorentzian WH geometries in non-minimally coupled scalar fields possessing torsion and boundary term in modified gravity. Much work is accessible in this hypothesis on different cosmological angles like the reconstruction of scalar fields, accelerated expanding universe, cosmological perturbation, Birkhoff's theorem, large-scale structure, thermodynamical laws validity, solar system limitations, etc \cite{12B}.

The existence of WH geometries is one of the most attractive subjects in contemporary cosmology. WHs are as the passage-like topological structures connecting two remote elements of the same universe or special universes through a shortcut called a tunnel or bridge. WHs, in general, are classified into two kinds of structures, specifically, the static WHs geometries, and the dynamical WH structures \cite{48}. The presence of a WH is an ultimate consequence of a solution of Einstein Field Equations which contain a non-trivial organized linkage of scattered objects in spacetime. Despite the speculative reality of the WHs, these associations stay predictable with the GR. Significant distances like billion light-years or more; brief distances comprising not many meters; various universes; time-dependent diversified points in the Universe might be related to utilizing these WHs. The blend of space and time into just a solitary spacetime continuum, as anticipated in the exceptional hypothesis of special relativity, may permit one hypothetically to go across space and time through a WH, utilizing some particular right conditions.

Even though the WHs as a consequence of the Schwarzschild solution are not traversable in both of the directions, yet their presence stirred the associated investigations to consider the prospect of traversable WHs shaped by holding the WH throat open in the presence of exotic matter. Lorentzian WHs \cite{Whee}, WHs producing a foam geometry within a general relativistic spacetime manifold uncovered by Lorentzian manifold, and  Euclidean WHs are also examples of non-traversable WHs. The traversable WH solution was first presented by Morris and Thorne \cite{Mor}. This idea was very not quite the same as those formerly hypothesized by Einstein and Nathan Rosen in 1935 who accomplished an answer notable today as the Einstein-Rosen bridge \cite{Ein}. Also it was dissimilar to the charge-possessing infinitesimal WHs as told by Wheeler \cite{Whee} and to the WHs allowing the two-path traffic of the human being like objects. Notwithstanding their skeptical presence of a WH similar to this, the investigation of Moris and Thorne has opened different opportunities for additional productive and momentous exploration which incorporates the far-reaching investigation of vital crucial properties of WHs \cite{Hoch1}-\cite{Ovgun:2020yuv}, utilizing them as time-machine \cite{Moris}-\cite{Frol}, the connection issues concerning causality infringement \cite{Kim}-\cite{Haw1}, and the constituted quantum or Plank-scale WHs \cite{Visser}-\cite{Hoch4}. Moreover, \"Ovg\"un et al. constructed an exact wormhole solution in bumblebee gravity for investigating the consequences of spontaneous Lorentz violation \cite{Ovgun:2018xys}.

It merits referencing here that it is the extraordinary exotic matter which is liable for the presence of WHs in GR. This exotic matter includes stress-energy tensor ${T}_{\mu \nu}$ which turns into the explanation behind the infringement of the null energy condition (NEC). It is worth mentioning here that NEC has the illustration of ${T}_{\mu \nu}k^{\mu}k^{\nu}$ with $k^{\mu}$ representing the presence of null vector. Presently, this has turned out to be an exciting situation for the WH contemporary physics to find the sustainable circumstances for the exotic spacetimes through the incorporation of non-exotic matter sources. There exist several candidates indicating the references for higher-dimensional WH geometries such as Gauss-Bonnet theory of gravity \cite{Bha}, WH solutions concerning brane \cite{Anchor}, as well as Brans-Dicke theory \cite{Nandi}. An effective systematic approach may be found in \cite{Garattini} comprising the WH geometries. Lobo and Oliveira \cite{Lobo3} built the WH solutions in $f(R)$ theory of gravity and presented exact solutions in the context of $f(R)$ gravity by employing an explicit shape-function and various expressions of the equation of state parameter. Static WH geometries in $f(R)$ gravity were investigated by Sharif and Zahra. \"Ovg\"un et al. studied the quasinormal modes and greybody factors of $f(R)$ gravity minimally coupled to a cloud of strings in $2+1$ Dimensions \cite{Ovgun:2018gwt}. Moreover, particle acceleration by static black holes in a Model of $f(R)$ gravity was investigated by Halilsoy et al.  \cite{Halilsoy:2015qta}.

For the growth of the WH geometries, the existence of some exotic fluid (a presumptuous matter form) is mandatory which disturbs the null energy constraint (NEC) in GR. The violation of NEC  is considered as the fundamental requirement for the existence of WH geometries. Discovering WH arrangements in GR has consistently been an incredible test for scientists. Although GR permits the presence of WHs, it is important to initially change the issue area by including some additional terms (since the ordinary fluid fulfills the energy limits and subsequently abuses the fundamental standards for the presence of the WH). These additional terms are liable for energy-bound infringement and henceforth allow the presence of WH in GR. In 1935, Einstein and Rosen \cite{49} talked about the numerical models of WHs in GR and they acquired the WH arrangements known as Lorentzian WHs or Schwarzschild WHs. In 1988, it was indicated \cite{50} that WHs could be enormous enough for humanoid voyagers and even grant time travel.

Referring here to the literature \cite{51,59} concerning the WH geometries, various writers have developed WH solutions by including various sorts of fascinating issues like quintom, scalar field models, noncommutative geometrical models, and electromagnetic field, and so on, and got distinctive intriguing and genuinely suitable outcomes. Some significant and intriguing outcomes concerning the steady WH arrangements without the incorporation of any exotic matter content are examined in \cite{60,61}. In a work, \cite{62}, the presence of the WH arrangements and their various properties in the modified $f(R, T)$ gravity has been examined. The coordinates in ``on a D-brane,  might be assumed as operators of noncommutative origin", this is perhaps the most fascinating part of the noncommutative geometrical aspects of the string hypothesis, giving a numerical method to investigate some significant ideas of quantum gravity \cite{63,64}. The noncommutative geometrical approach is a push to build a unified stage where one may assume the spacetime forces of gravitation as a joined type of frail and solid forces due to gravity. The non-commutativity approach has the significant element of supplanting point-like constructions with dispersed items and henceforth compares to the spacetime quantification, which is because of the commutator characterized by $[x^\zeta, x^\eta]=\iota\theta^{\zeta\eta}$, where $\theta^{\zeta\eta}$ is an anti-symmetric matrix of second order. This dispersal impact can be demonstrated by including Gaussian and Lorentzian distributions of insignificant length $\sqrt{\theta}$ rather than the Dirac delta function. The spherically symmetric, static particle-like gravitational source addressing the geometry of Gaussian distribution with non-commutative nature with the maximum mass M possesses.
While, regarding the Lorentzian dispersion, we can take the density capacity of the particle-like mass $M$. Here, the entire mass $M$ can be thought as a WH, a form of diffused unified object and, $\theta$ being the noncommutative parameter. The Gaussian matter distribution has been employed by Sushkov to model the phantom-energy supported WHs \cite{68}. Likewise, Nicolini and Spalluci \cite{69}, implemented this type of matter distribution to establish the physical implications of short-separated variations of non-commutating coordinates in exploring black holes.

Having enthused by the work in teleparallel gravity and other modified theories of gravity, here in our work,
spherically symmetric static WH geometries have been constructed under the modification of $f(\tau)$ gravity with matter coupling $T$. In Section II, the basic description of the mathematical formulation of $f(\tau,T)$ gravity along with the fundamental criteria for the existence of WH geometries, exclusive expressions for the energy density and stress profiles, and corresponding energy bounds are provided. In Section III,  the traversable WH solutions by incorporating the linear model of $f(\tau, T)=\alpha\tau(r)+\beta T+\phi$ for both the Gaussian and Lorentzian sources of non-commutative geometry have been provided. In Section IV, the stability of the Gaussian and Lorentzian WH geometries is discussed through graphical analysis. Section V gives the embedding diagrams of our traversable WH solutions of the Gaussian and Lorentzian sources. Section VI concludes our entire work.

\section{ $f(\tau,T)$ Gravity}
The modification of $f(\tau)$ gravity with matter coupling $T$ is defined with the following modified action
\begin{equation}
S_{A}=\int dx^{4} e\lbrace \frac{1}{2k^{2}} f(\tau,T)+\mathcal{L}_{(m)}\rbrace \label{1},
\end{equation}
where,  $e= det\left( e_{\mu}^{A}\right) =\sqrt{-g}$, $k^{2}=8\pi G =1$ and, $T= \delta_{\gamma}^{\varepsilon\alpha}\tau_{\varepsilon\alpha}^{\gamma}=[-\rho,+p_r,+p_t,+p_t]$.
$\mathcal{L}_{(m)}$ represents Lagrangian density. The Contorsion, Torsion, and super potential are the key elements of this modified gravity, and are expressed as
\begin{eqnarray}
K^{\varepsilon\upsilon}_{\lambda}&&=-\frac{1}{2}\left(\tau^{\varepsilon\upsilon}{}_{\lambda}-\tau^{\upsilon\varepsilon}{}_{\rho}-\tau_{\lambda} {}^{\varepsilon\upsilon}\right),\;\;
\tau^{\lambda}_{\varepsilon\upsilon}=e_{\vartheta}{}^{\lambda}(\partial_{\varepsilon}e^\vartheta{}_\upsilon-\partial_{\upsilon}
e^\vartheta{}_{\varepsilon}),\;\;
S_{\lambda}{}^{\varepsilon\upsilon}=\frac{1}{2}\left(K^{\varepsilon\upsilon}{}_{\lambda}+
\delta^{\varepsilon}{}_{\lambda}{\tau^{\gamma\varepsilon}{}_{\gamma}}-\delta^{\upsilon}{}_{\lambda} {\tau^{\gamma\varepsilon}{}_{\gamma}}\right).\label{2}
\end{eqnarray}
In Eq. (\ref{2}), the term $\tau$ is defined by the relation, i.e.,$\tau=\tau^{\lambda}{}_{\kappa\upsilon}S_{\lambda}{}^{\kappa\upsilon}$.

The spherically symmetric geometry is chosen for this study, the line element of which is expressed as:
\begin{equation} \label{3}
ds^{2}=-e^{a(r)}dt^{2}+e^{b(r)} dr^{2}+r^{2} (d\theta^{2}+\sin^{2}\theta d\phi^{2}) ,
\end{equation}
where $a(r)=2\varpi(r)$ and $b(r)=log\left(1-\frac{\epsilon _s(r)}{r}\right)^{-1}$. Here, $\varpi(r)$ and $\epsilon _s(r)$ represent the red-shift, and the WH shape functions, respectively. Both the functions have some necessary conditions, which are summarized as\\
$\bullet$ Redshift function should be positive and free from any horizon.\\
$\bullet$ In this study, redshift function is kept constant, i.e., $\varpi^{'}=0.$\\
$\bullet$ WH shape-function must be positive with increasing behavior.\\
$\bullet$ WH throat must exist, i.e., $\epsilon _s(r_0)=r_0$.\\
$\bullet$ The ratio of WH shape function and radial coordinate must approach to zero as radial coordinate approaches to infinity, i.e., $\frac{\epsilon _s}{r}\rightarrow0$.\\
$\bullet$ WH shape function should fulfill the flaring out condition, i.e., $\epsilon _{s}^{'}\mid_{r=r_0}<1$.\\
$\bullet$ The critical constraint of $b_{\epsilon}'(r_0)<1$ must be satisfied for the existence of WH structure.\\
Now, the diagonal tetrad is calculated as:
\begin{equation}\label{4}
e_{\gamma}^{\eta} =\Big(e^{\frac{a(r)}{2}},e^{\frac{b(r)}{2}},r,r\sin\theta\Big).
\end{equation}
The determinant of $e_{\gamma}^{\eta}$ is calculated as $e=e^{a(r)+b(r)}r^{2}\sin\theta$. The anisotropic energy momentum tensor is expressed as:
\begin{equation} \label{5}
{T}_{\alpha \omega}=(\rho+p_{t})u_{\alpha}u_{\omega}-p_{t}g_{\alpha \omega}+(p_{r}-p_{t})v_{\alpha}v_{\omega}.
\end{equation}
where $u_{\alpha}=e^{\frac{a}{2}}\delta_{\alpha}^{0}$ and $v_{\alpha}=e^{\frac{b}{2}}\delta_{\alpha}^{1}$. Here $\rho$, $p_{r}$, and $p_{t}$ represent the energy density, radial and tangential components of pressure, respectively. From Eq. (\ref{1}), we can get the generalized field equations of $f(\tau,T)$ gravity by using the energy momentum tensor as:
\begin{equation}\label{6}
\Big[e^{-1}\partial_{\varepsilon}(e e_{a}^{\alpha}S_{\alpha}^{\omega\varepsilon})+e_{a}^{\alpha}\tau^{\varepsilon}_{a\alpha}S_{\varepsilon}^{a \omega}\Big]f_{\tau}+e_{a}^{\alpha}S_{\alpha}^{\omega\varepsilon}(f_{\tau\tau}\partial_{\varepsilon}\tau+f_{\tau T}\partial_{\varepsilon}T)+\frac{e_{a}^{\omega}f}{4}-
(\frac{e_{a}^{\alpha}T_{\alpha}^\omega+p_{t}e_{a}^{\omega}}{2})f_{T}=\frac{e_{a}^{\alpha}{T}_{\alpha}^{\omega}}{4}.
\end{equation}
The valid expression for $\tau$, for a WH space-time takes the following form:
\begin{equation}
\tau(r)=\frac{2 e^{-b (r)}}{r^{2}}\label{7}.
\end{equation}
Here, we use suitable linear model for $f(\tau,T)$ gravity with diagonal tetrad, as follows:
\begin{equation}
f(\tau,T)=\alpha\tau(r)+\beta T+\psi\label{8}.
\end{equation}
where $\alpha,\;\beta$ and $\psi$ are constants. By plugging Eq. (\ref{3}), Eq. (\ref{5}), and Eqs. (\ref{7}-\ref{8}) in Eq. (\ref{6}), we get the following field equations for $f(\tau,T)$ gravity:
\begin{eqnarray}
\rho &=&\frac{-(\beta +2)r^3\psi+\alpha(\beta -4)r\epsilon _s'(r)+\alpha\beta\epsilon _s(r)}{4\left(\beta ^2+\beta -2\right)r^3},\label{9}\\
p_r&=&\frac{(\beta +2)r^3\psi+3\alpha\beta r\epsilon _s'(r)+\alpha(4-5 \beta)\epsilon _s(r)}{4 \left(\beta ^2+\beta -2\right)r^3},\label{10}\\
p_t&=&\frac{(\beta +2)r\left(r^2\psi+\alpha\epsilon _s'(r)\right)+\alpha(\beta -2)\epsilon _s(r)}{4\left(\beta ^2+\beta -2\right)r^3}.\label{11}
\end{eqnarray}

Now, we discuss  the energy conditions (ECs), i.e., null energy condition $(NEC)$, weak energy condition $(WEC)$, strong energy condition $(SEC)$ and dominant energy condition $(DEC)$, and are read as
\begin{eqnarray}
\nonumber NEC&:&\rho+p_r\geq 0,~~~~~~~~~~\rho+p_t\geq 0,\\
\nonumber WEC&:&\rho\geq 0,~~~~~~~~~~~~~~~~~\rho+p_r\geq 0,~~~~~~~~~~\rho+p_t\geq 0,\\
\nonumber SEC&:&\rho+p_r\geq 0,~~~~~~\rho+p_t\geq 0,~~~~~~~~~~\rho+p_r+2p_t\geq 0,\\
\nonumber DEC&:&\rho\geq 0,~~~~~~~~~~~~~~~~~\rho-|p_r|\geq 0,~~~~~~~~\rho-|p_t|\geq 0.
\end{eqnarray}
Theses ECs for $f(\tau,T)$ gravity are satisfied by the normal matter because of positive density and positive pressure. To discuss the WH construction, we shall check the behavior of the ECs, as the $NEC$ violation is the necessary requirement for the existence of exotic matter.
\section{The study of traversable wormhole}

In this section, we shall explore different features of WH geometry. In order to calculate the shape function of WH, we consider the smearing effects by using the noncommutative geometry. For this purpose, we shall plug two kinds (Gaussian and Lorentzian) of non-commutative geometries in this study. The Gaussian and Lorentzian sources of energy density are expressed as:
\begin{equation}
\rho=\frac{M e^{-\frac{r^2}{4 \theta }}}{8 \pi ^{3/2} \theta ^{3/2}},\;\;\;\;\;\rho=\frac{\sqrt{\theta } M}{\pi ^2 \left(\theta +r^2\right)^2}\label{12}.
\end{equation}
where $M$ is the total particle mass, $\theta$ is the direct non-commutative parameter for Lorentzian and Gaussian non-commutative geometries. By comparing Eq. (\ref{9}) and Eq. (\ref{12}), we get the following differential equations
\begin{eqnarray}
\frac{-(\beta +2) r^3 \psi +\alpha  (\beta -4) r \epsilon _s'(r)+\alpha  \beta  \epsilon _s(r)}{4 \left(\beta ^2+\beta -2\right) r^3}&&=\frac{M e^{-\frac{r^2}{4 \theta }}}{8 \pi ^{3/2} \theta ^{3/2}},\label{13}\\
\frac{-(\beta +2) r^3 \psi +\alpha  (\beta -4) r \epsilon _s'(r)+\alpha  \beta  \epsilon _s(r)}{4 \left(\beta ^2+\beta -2\right) r^3}&&=\frac{\sqrt{\theta } M}{\pi ^2 \left(\theta +r^2\right)^2}.\label{14}
\end{eqnarray}
On solving Eq. (\ref{13}), we get the following shape function of WH for the Gaussian source of the non-commutative geometry
\begin{equation}
\epsilon _s(r)=((\beta -4) (-r))^{\frac{\beta }{4-\beta }}\left(C_{1}-\frac{(\beta +2) r^4 ((\beta -4) (-r))^{\frac{4}{\beta -4}} \left(\frac{\pi ^{3/2} (\beta -4) \psi }{\beta -3}-\frac{(\beta -1) M\times ExpIntegralE \left(-\frac{2}{\beta -4}-1,\frac{r^2}{4 \theta }\right)}{\theta ^{3/2}}\right)}{4 \pi ^{3/2} \alpha }\right)\label{15}.
\end{equation}
where $``ExpIntegralE"$ is a special function, and it can be written as $E_{-\frac{2}{\beta -4}-1}\left(\frac{r^2}{4 \theta }\right)$ and $C_{1}$ is a constant of integration. Now, by plugging Eq. (\ref{15}) in Eqs. (\ref{9}-\ref{11}), we get energy density and components of pressure for Gaussian distribution as:
\begin{eqnarray}
\rho &=&\frac{M e^{-\frac{r^2}{4 \theta }}}{8 \pi ^{3/2} \theta ^{3/2}},\label{16}\\
p_r&=&\frac{1}{32 \left(\beta ^2+\beta -2\right) r^3}\times\bigg(4\bigg(\frac{6 \alpha  \beta ^2 C_{1} ((\beta -4) (-r))^{-\frac{4}{\beta -4}}}{(\beta -4)^2 r}-2 \alpha  (5 \beta -4) C_{1} ((\beta -4) (-r))^{\frac{\beta }{4-\beta }}\nonumber\\&+&\frac{(\beta -1) (\beta +2) r^3 \left(\frac{(\beta +1) M e^{-\frac{r^2}{4 \theta }}}{\pi ^{3/2} \theta ^{3/2}}+4 \psi \right)}{\beta -3} \bigg)+\frac{2 (\beta -1)^2 \left(\beta ^2-4\right) M r^5 E_{-2-\frac{2}{\beta -4}}\left(\frac{r^2}{4 \theta }\right)}{\pi ^{3/2} (\beta -4) (\beta -3) \theta ^{5/2}}\bigg),\label{17}\\
p_t&=&\frac{\frac{4 (\beta -1) \left(\frac{16 \alpha  C_{1} ((\beta -4) (-r))^{-\frac{4}{\beta -4}}}{(\beta -4)^2}+\frac{(\beta +2) r^4 \left(\frac{(\beta +1) M e^{-\frac{r^2}{4 \theta }}}{\pi ^{3/2} \theta ^{3/2}}+4 \psi \right)}{\beta -3}\right)}{r}+\frac{2 \left(\beta ^3-3 \beta +2\right) M r^5 E_{-2-\frac{2}{\beta -4}}\left(\frac{r^2}{4 \theta }\right)}{\pi ^{3/2} (\beta -4) (\beta -3) \theta ^{5/2}}}{32 \left(\beta ^2+\beta -2\right) r^3}.\label{18}
\end{eqnarray}
\begin{figure}
\centering \epsfig{file=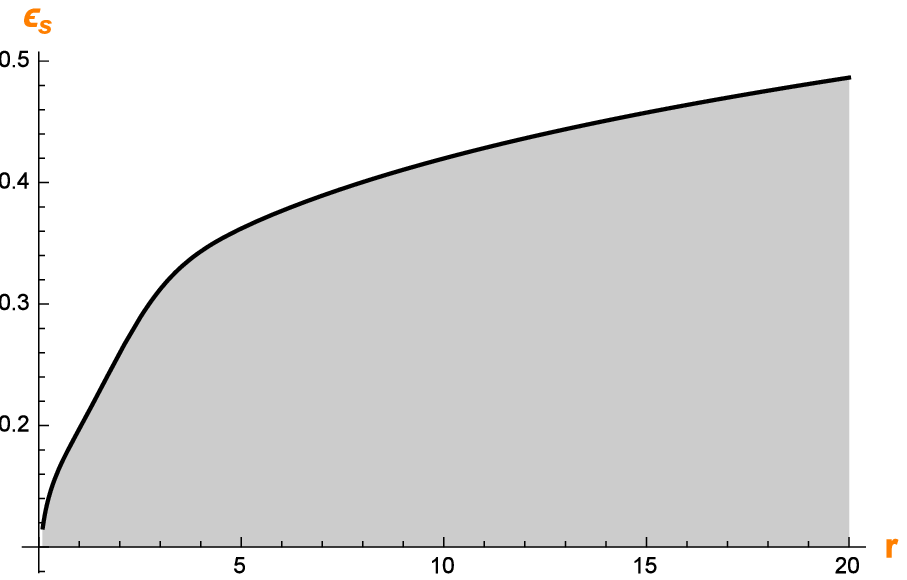, width=.32\linewidth,
height=2.02in}\epsfig{file=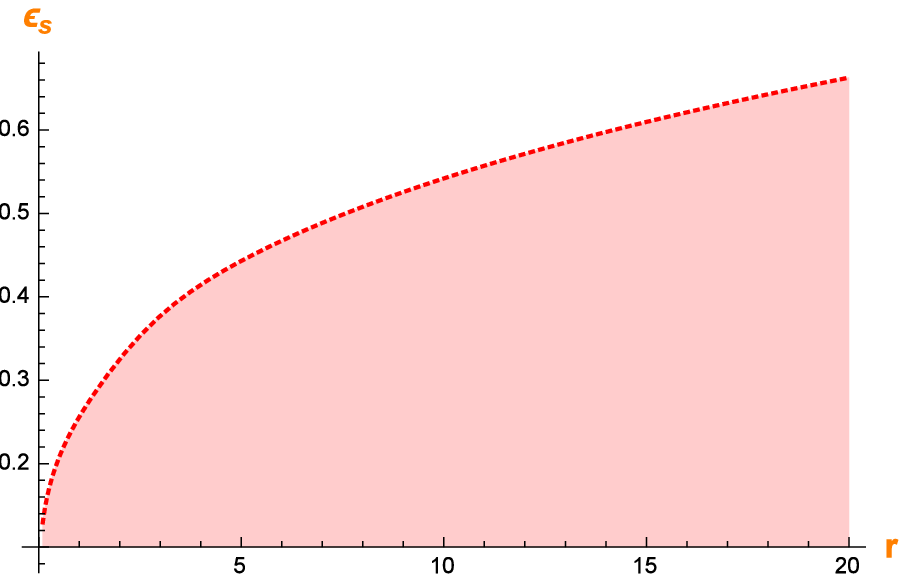, width=.32\linewidth,
height=2.02in}\epsfig{file=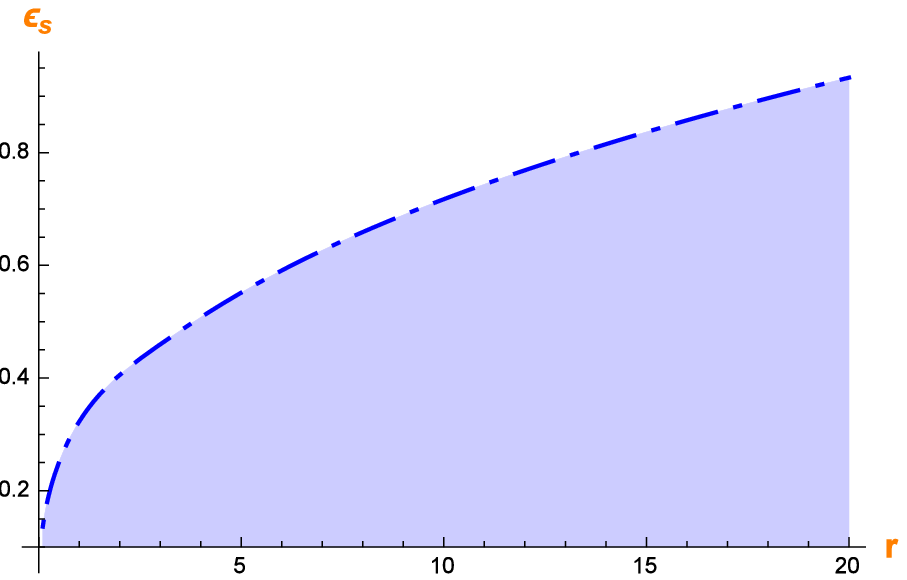, width=.32\linewidth,
height=2.02in}
\centering \epsfig{file=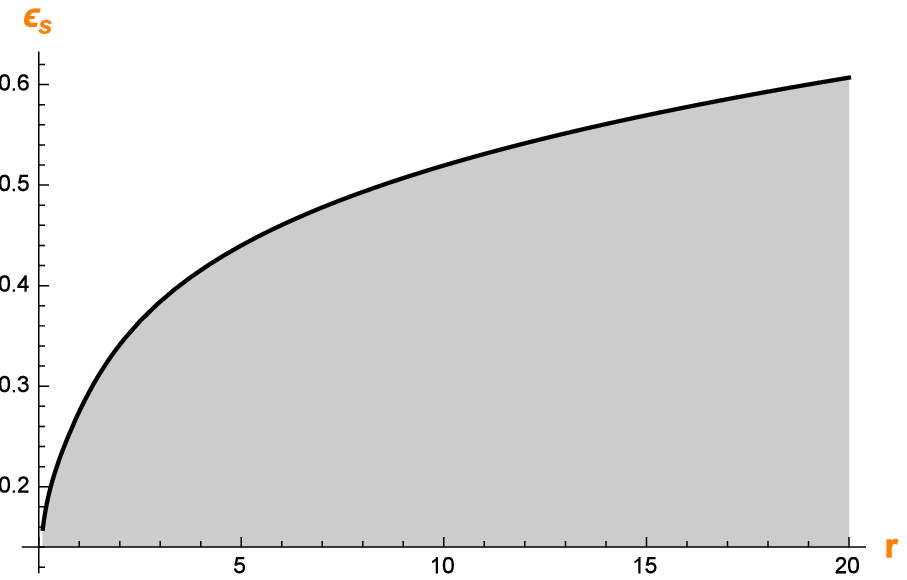, width=.32\linewidth,
height=2.02in}\epsfig{file=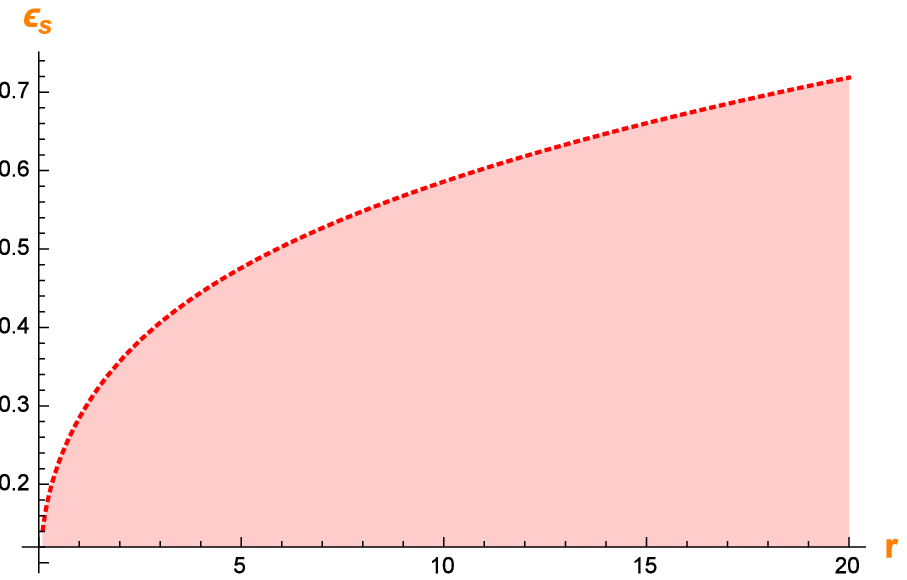, width=.32\linewidth,
height=2.02in}\epsfig{file=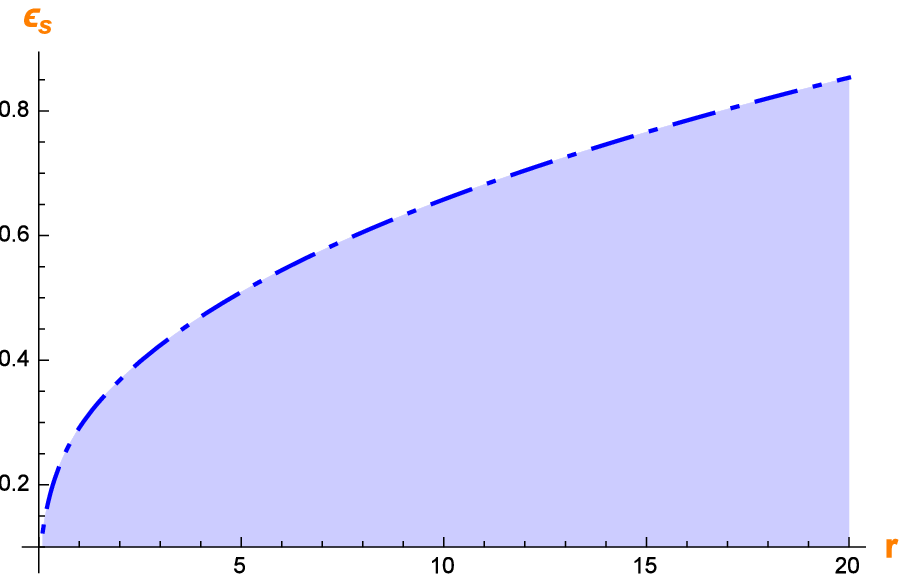, width=.32\linewidth,
height=2.02in} \caption{\label{fig1} shows the required behavior of $\epsilon _s(r)$.}
\end{figure}

\begin{figure}
\centering \epsfig{file=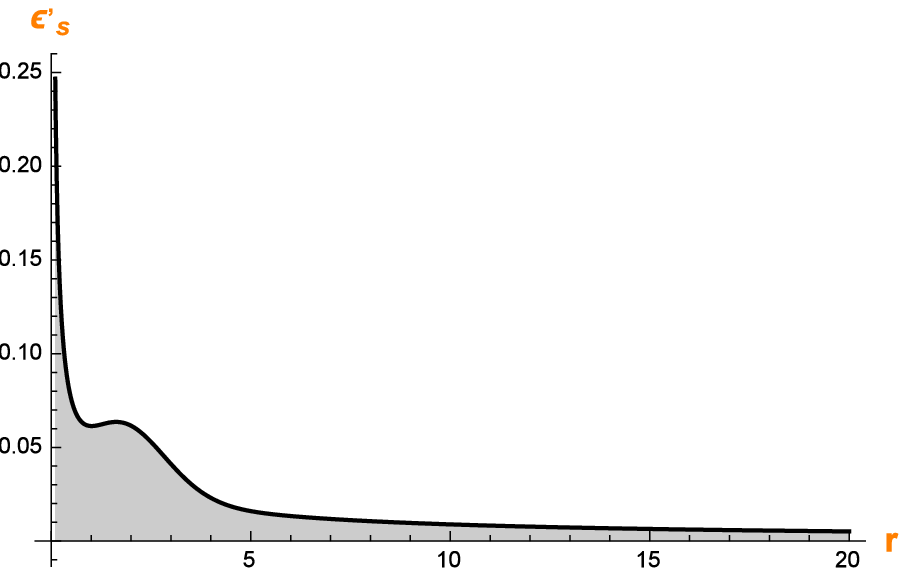, width=.32\linewidth,
height=2.02in}\epsfig{file=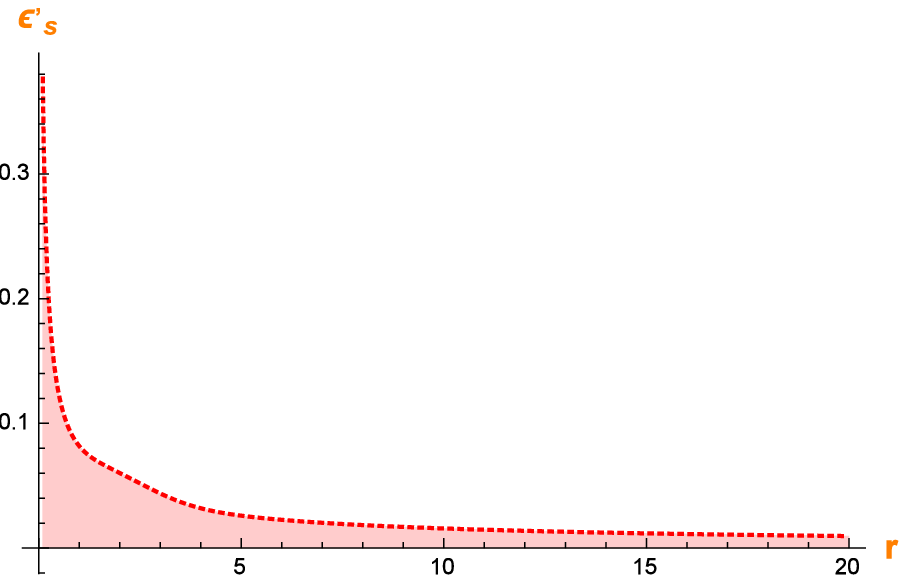, width=.32\linewidth,
height=2.02in}\epsfig{file=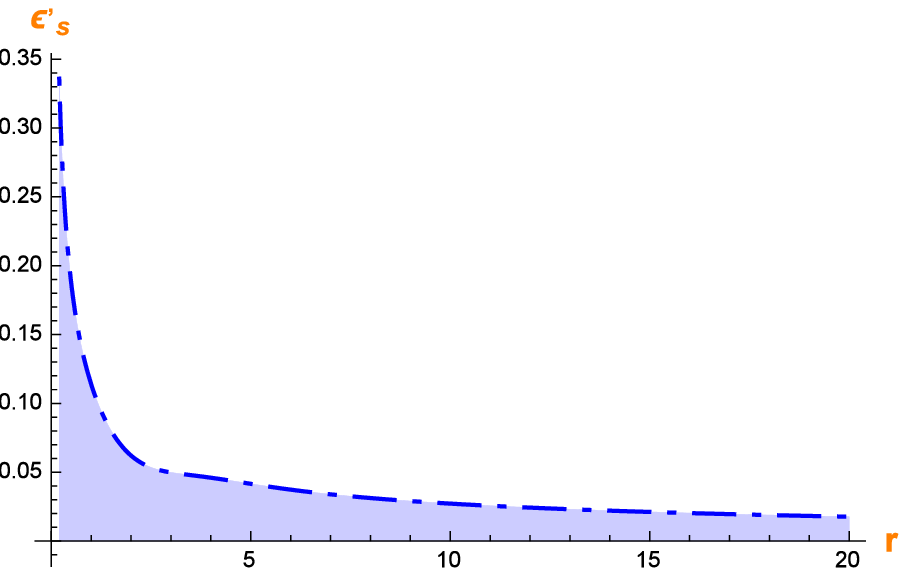, width=.32\linewidth,
height=2.02in}
\centering \epsfig{file=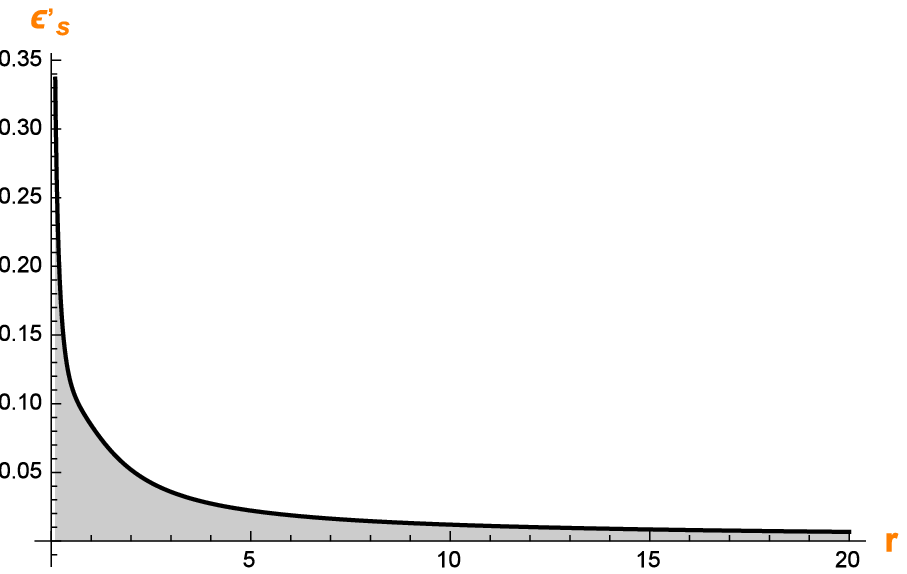, width=.32\linewidth,
height=2.02in}\epsfig{file=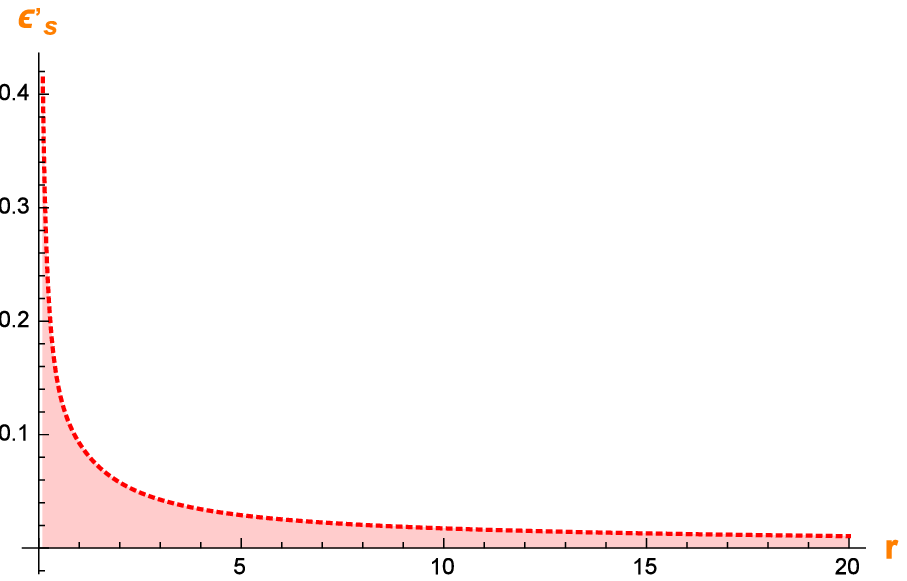, width=.32\linewidth,
height=2.02in}\epsfig{file=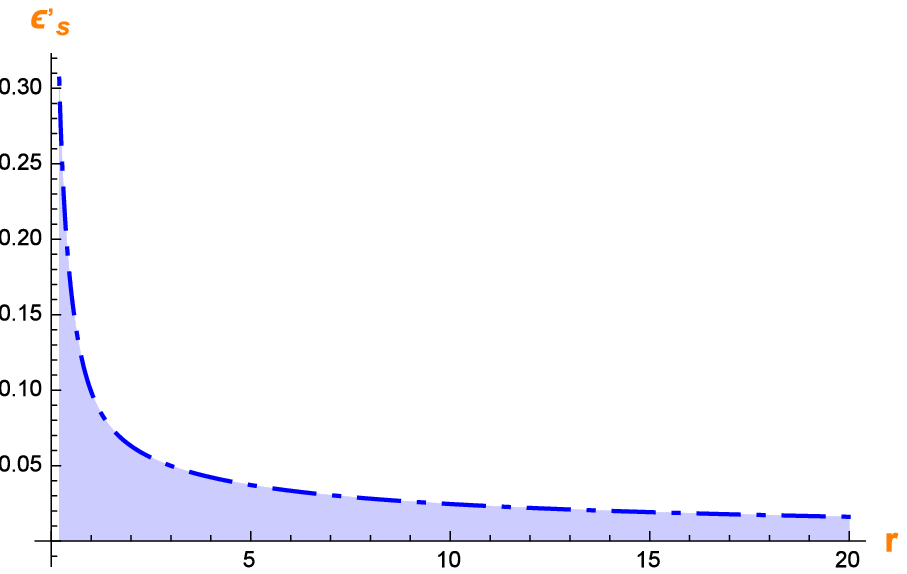, width=.32\linewidth,
height=2.02in} \caption{\label{fig2} shows the required behavior of $\frac{d\epsilon _s}{dr}$.}
\end{figure}

\begin{figure}
\centering \epsfig{file=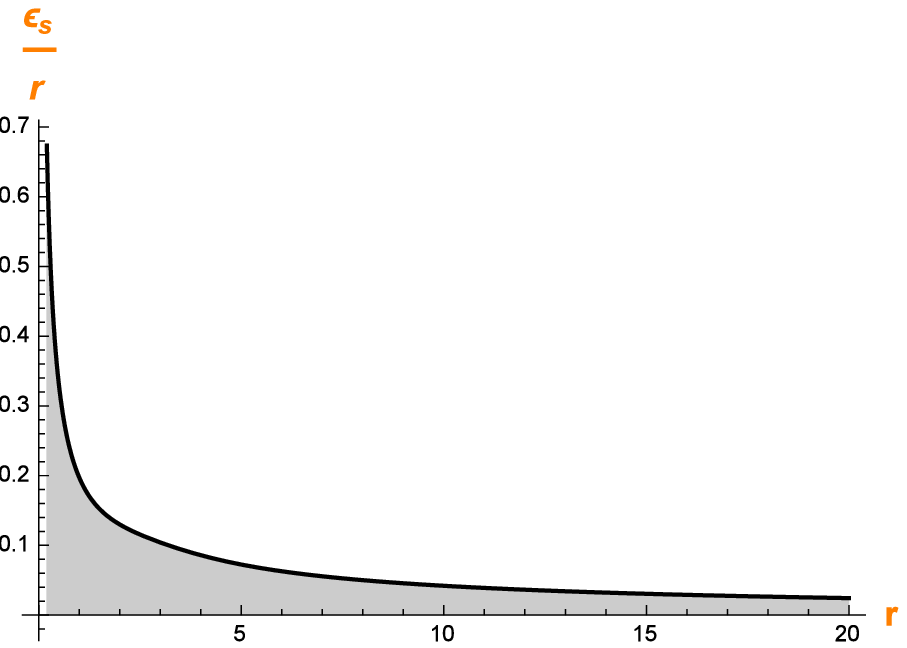, width=.32\linewidth,
height=2.02in}\epsfig{file=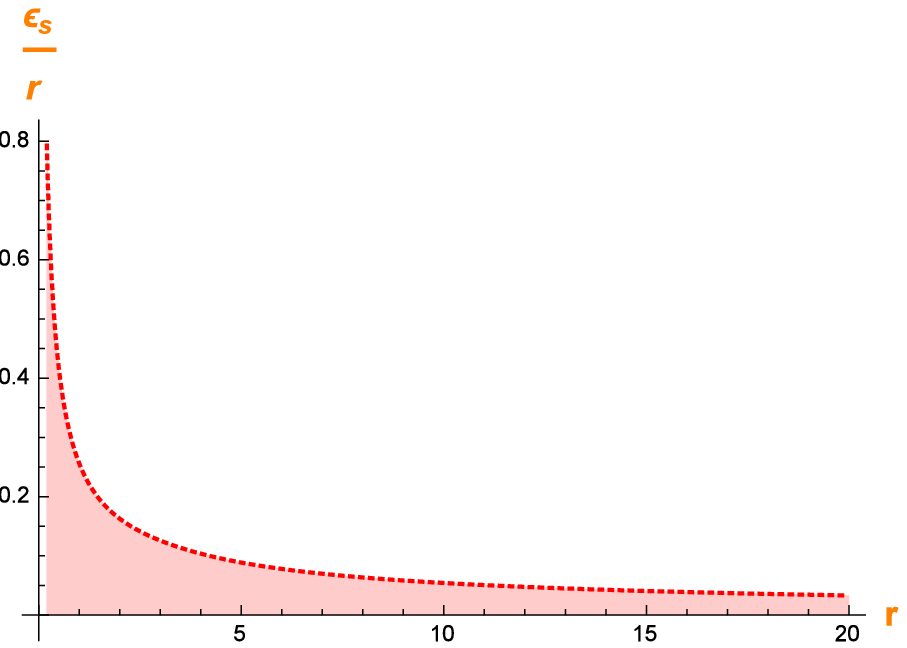, width=.32\linewidth,
height=2.02in}\epsfig{file=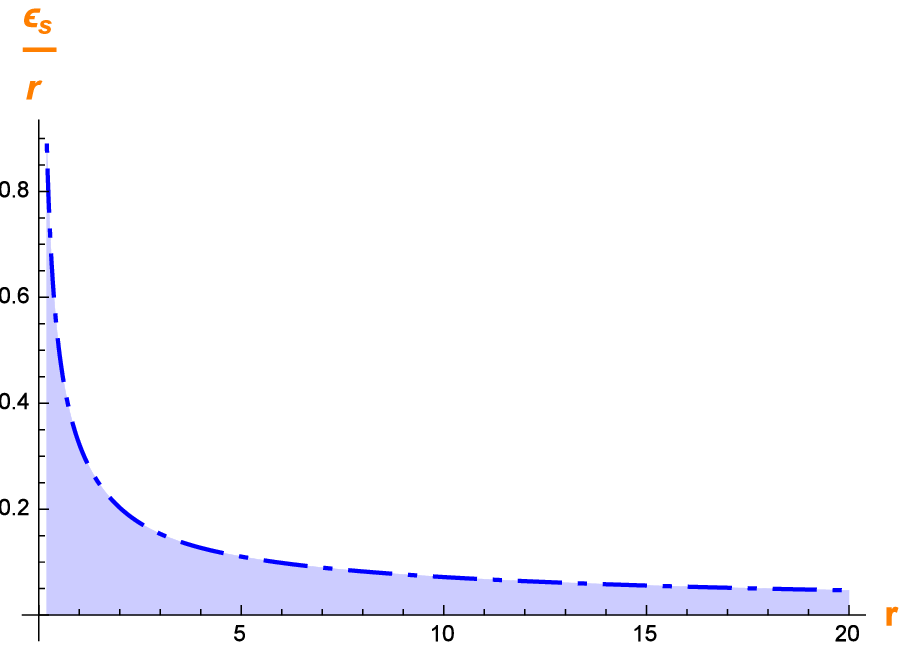, width=.32\linewidth,
height=2.02in}
\centering \epsfig{file=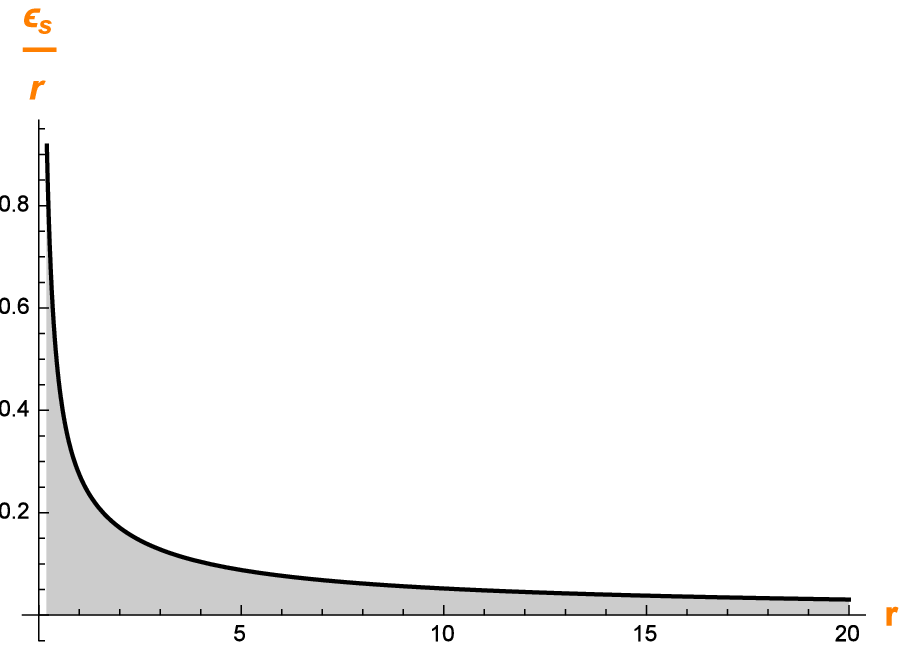, width=.32\linewidth,
height=2.02in}\epsfig{file=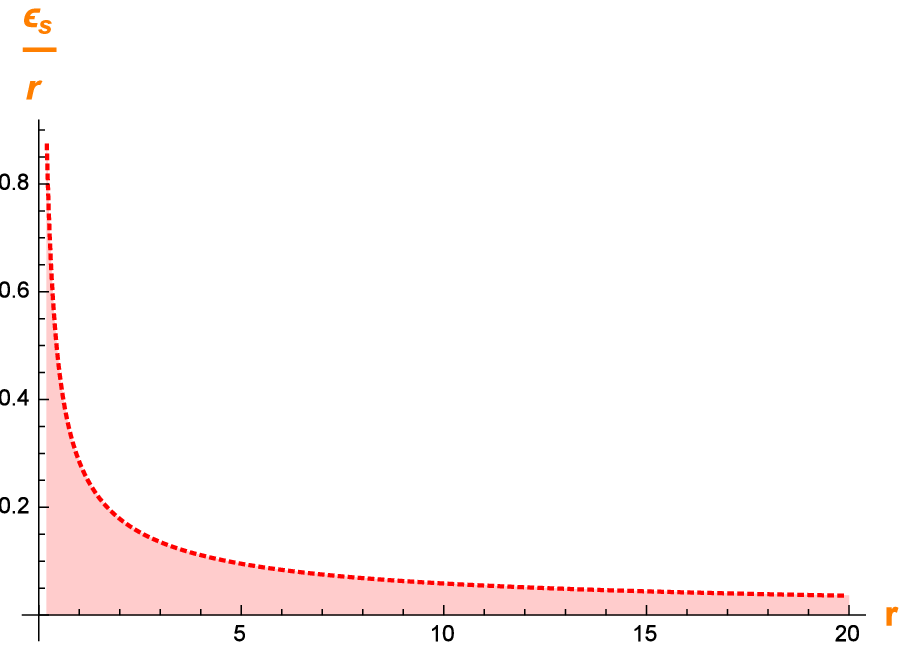, width=.32\linewidth,
height=2.02in}\epsfig{file=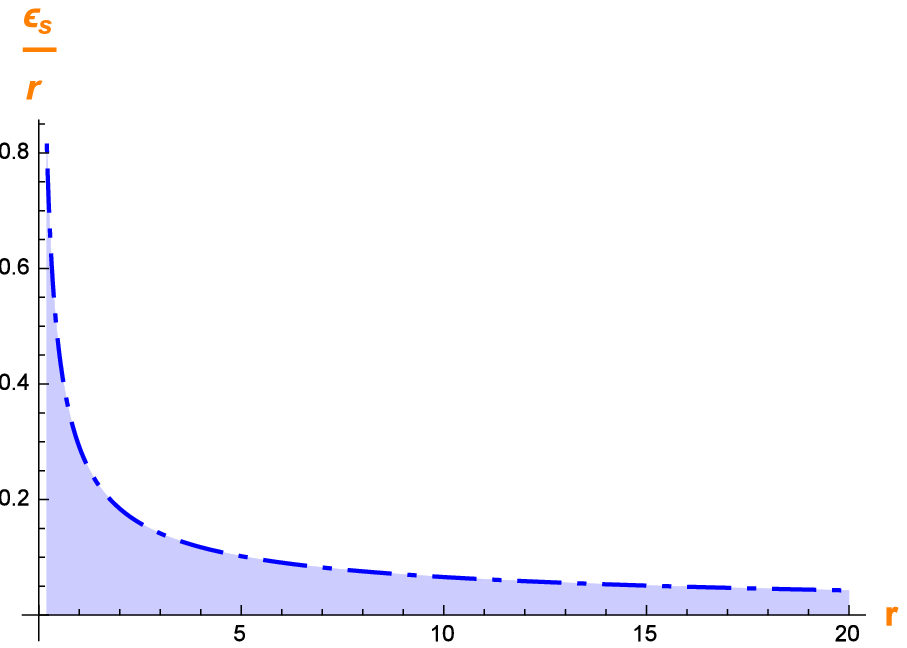, width=.32\linewidth,
height=2.02in} \caption{\label{fig3} shows the required behavior of $\frac{\epsilon _s}{r}$.}
\end{figure}

\begin{figure}
\centering \epsfig{file=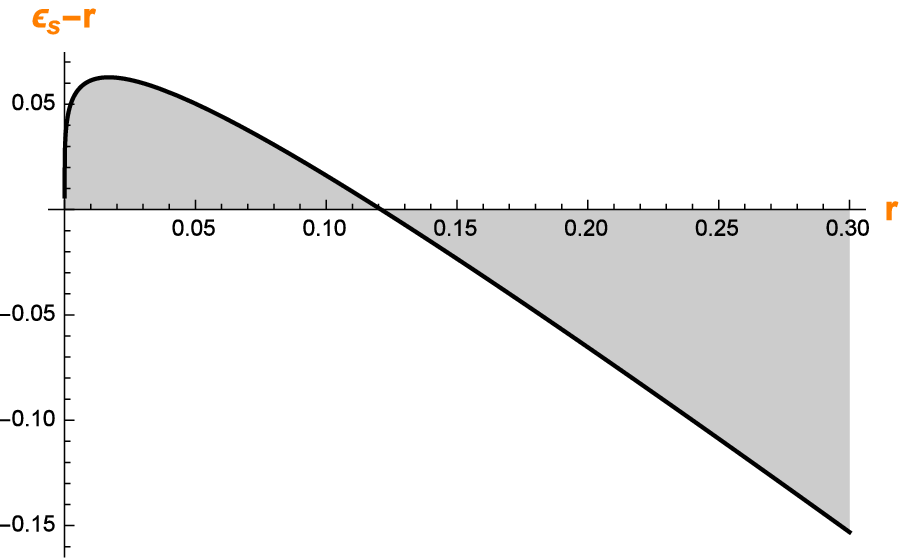, width=.32\linewidth,
height=2.02in}\epsfig{file=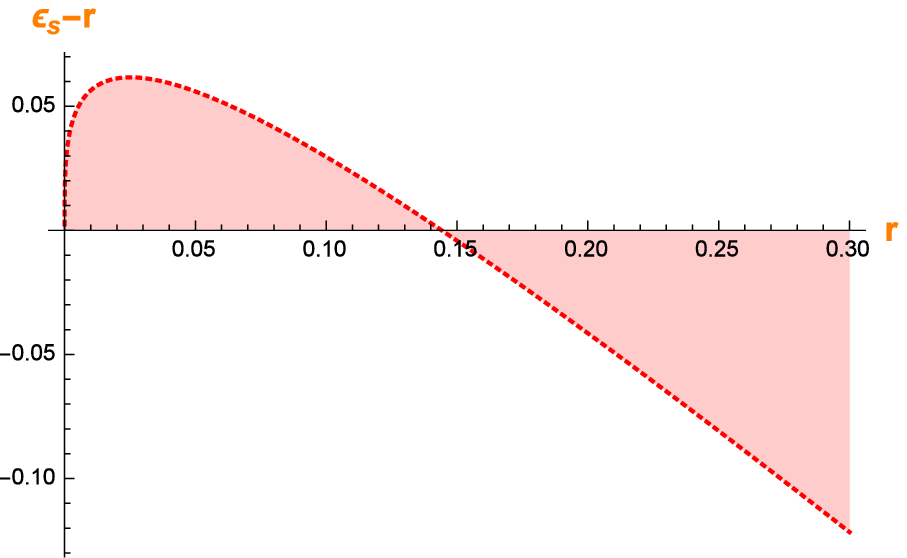, width=.32\linewidth,
height=2.02in}\epsfig{file=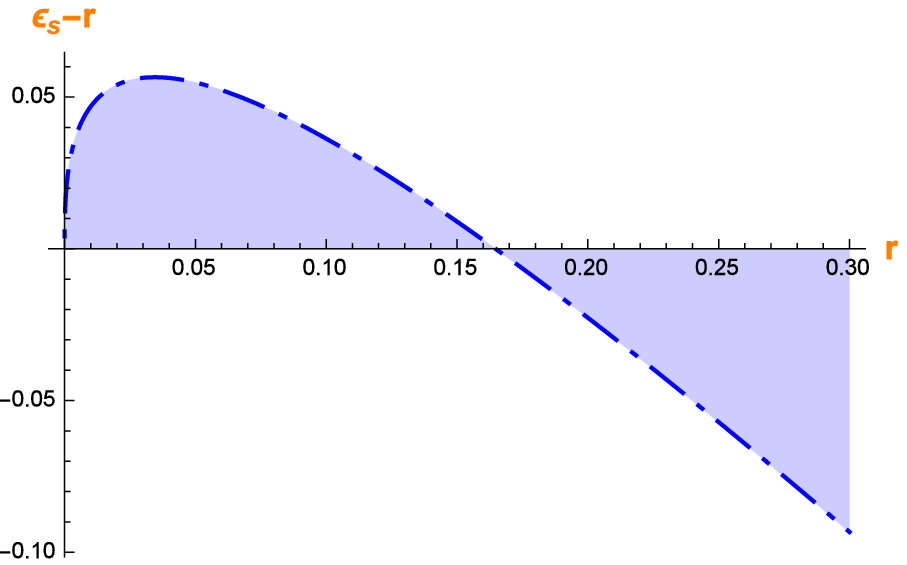, width=.32\linewidth,
height=2.02in}
\centering \epsfig{file=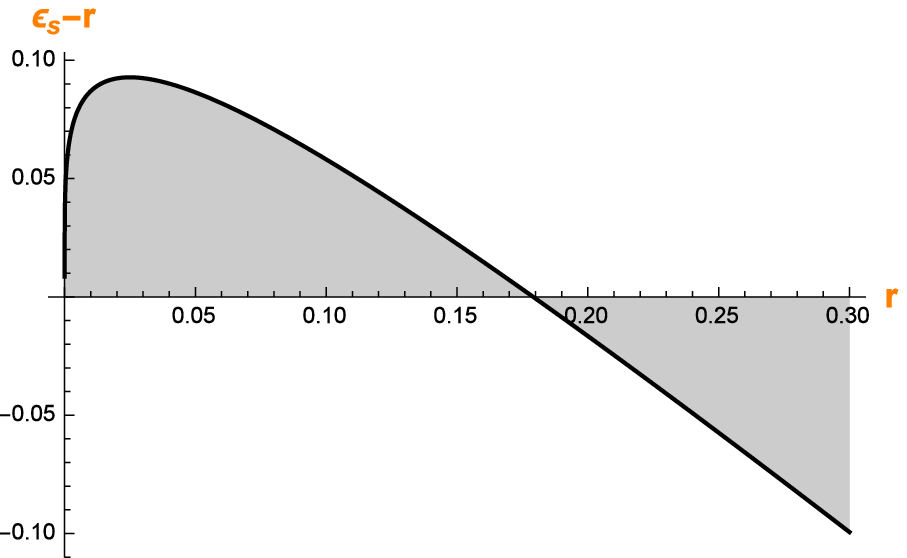, width=.32\linewidth,
height=2.02in}\epsfig{file=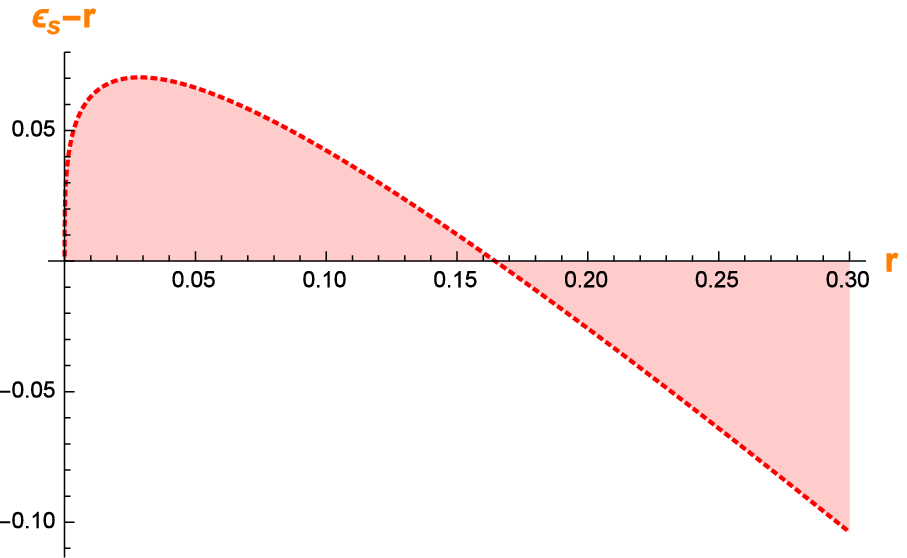, width=.32\linewidth,
height=2.02in}\epsfig{file=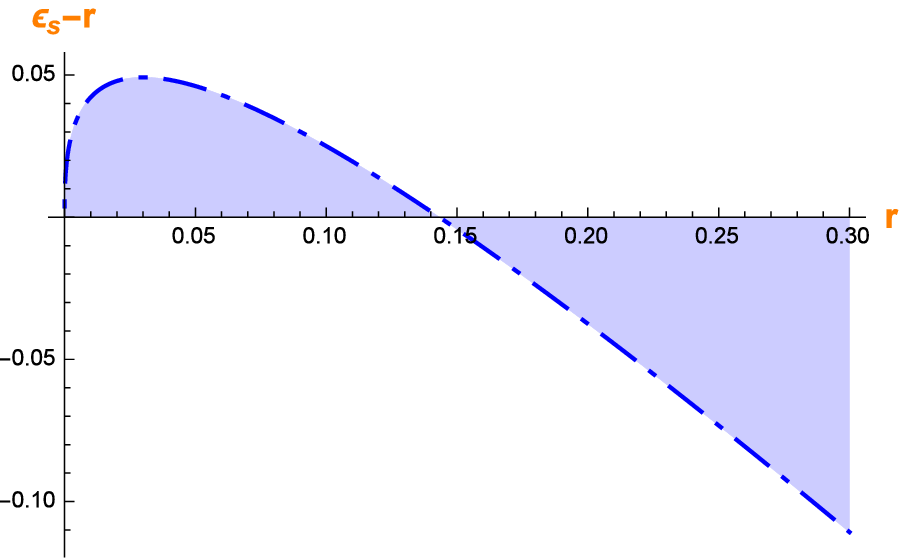, width=.32\linewidth,
height=2.02in} \caption{\label{fig4} shows the required behavior of $\epsilon _{s}-r$.}
\end{figure}
The graphical behavior of shape function $\epsilon _s(r)$ is provided in Fig. (\ref{1}) for both the Gaussian and the Lorentzian distributions. In the first row, the $\epsilon _s(r)$ for the Gaussian framework are described with three different values of the matter coupling parameter, i.e., $\beta=0.70,\;0.90,\;1.10$ by left, middle, and right parts respectively. The second row represents the graphical analysis for the Lorentzian framework for $\beta=0.70,\;0.90,\;\&\;1.10$ by left, middle, and right parts respectively. It can be verified from Fig. (\ref{1}) that $\epsilon _s(r)$ is regularly increasing with positive behavior. The increasing behavior of the shape function shows that our calculated shape functions in both Gaussian and Lorentzian distributions are well-fitted for the WH study. The derivative of shape function $\epsilon _s(r)$ can be seen in Fig. (\ref{2}) for both distributions. In first row the $\frac{d\epsilon _s}{dr}$ under Gaussian framework is shown for $\beta=0.70,\;0.90,\;\&\;1.10$ by left, middle, and right portions, respectively. $\frac{d\epsilon _s}{dr}$ for the Lorentzian framework is provided in the second row of Fig. (\ref{2}) for $\beta=0.70,\;0.90,\;\&\;1.10$ by left, middle, and right portions, respectively. The required behavior $\frac{d\epsilon _s}{dr}<1$ may be verified from Fig. (\ref{2}). The existence of the constraint $\frac{d\epsilon _s}{dr}<1$ depicts that our calculated results in both the distributions satisfy the flaring out condition of WH.\\

The ratio $\frac{\epsilon _s}{r}$ vanishes as $r$ approaches to infinity, i.e., $\frac{\epsilon _s}{r}\rightarrow 0$ as $r\rightarrow\infty$ in both cases, which can be verified from Fig. (\ref{3}). The acquiring of the expression $\frac{\epsilon _s}{r}\rightarrow 0$ as $r\rightarrow\infty$ for $\beta=0.70,\;0.90,\;\&\;1.10$ by left, middle, and right portions respectively validates the flatness property of the space-time. This property exhibits that WH space-time should be flat under both Gaussian and Lorentzian distributions in this scenario. The flatness condition is indispensable in WH study the fulfillment of which depicts the overwhelming role of the noncommutative geometry in the ongoing WH study. The WH throats are calculated via $\epsilon _{s}-r$ for both Gaussian and Lorentzian distributions. From the first row of Fig. (\ref{4}), it can be perceived that we get different values of WH throats against the different values of parameter $\beta$,i.e., $\beta=0.70,\;0.90,\;\&\;1.10$ by left, middle, and right parts respectively. In the Gaussian framework, the WH throats are calculated as $r_{0}=0.120$ for $\beta=0.70$, $r_{0}=0.145$ for $\beta=0.90$, and $r_{0}=0.165$ for $\beta=1.10$. In the Lorentzian distribution, the WH throats are calculated as $r_{0}=0.180$ for $\beta=0.70$, $r_{0}=0.165$ for $\beta=0.90$, and $r_{0}=0.145$ for $\beta=1.10$, these throats can be verified from the second row of Fig. (\ref{4}). The different values of WH throat show the critical impact of parameter $\beta$ in the current scenario. The different values of parameter $\beta$ provide the different WH throat locations. All the properties for both the cases are provided in Tabs. (\ref{I}-\ref{II})\\
\begin{center}
\begin{table}
\caption{\label{I}{Detailed summary of WH properties for Gaussian non-commutative distribution under $\alpha =0.5$, $C_{1}=0.2$, $\theta =0.9$, $ M=0.5$, and $\psi =3.036\times10^{-34}$}}
\begin{tabular}{|c|c|c|c|c|c|c|c|c|}
    \hline
\multicolumn{4}{|c|}{Gaussian Noncommutative Distribution}\\
    \hline
$Parameter/Expressions$    & $\beta=0.70$          &$\beta=0.90$  &$\beta=1.10$\\
\hline
$\epsilon _s(r)$            & $\epsilon _s(r)>0$ in $0.1\leq r\leq20$   &$\epsilon _s(r)>0$ in $0.1\leq r\leq20$  &$\epsilon _s(r)>0$ in $0.1\leq r\leq20$\\

$\frac{d\epsilon _s}{dr}$   & $\frac{d\epsilon _s}{dr}\mid_{r_{0}}<1$ in $0.1\leq r\leq20$    & $\frac{d\epsilon _s}{dr}\mid_{r_{0}}<1$ in $0.1\leq r\leq20$  &$\frac{d\epsilon _s}{dr}\mid_{r_{0}}<1$ in $0.1\leq r\leq20$\\

$\frac{\epsilon _s}{r}$          & $\frac{\epsilon _s}{r}\rightarrow 0$ as $r\rightarrow\infty$  &$\frac{\epsilon _s}{r}\rightarrow 0$ as $r\rightarrow\infty$  &$\frac{\epsilon _s}{r}\rightarrow 0$ as $r\rightarrow\infty$\\

$\epsilon _{s}-r$                & $r_{0}=0.120$  &$r_{0}=0.145$  &$r_{0}=0.165$\\
\hline
\end{tabular}
\end{table}
\end{center}

\begin{center}
\begin{table}
\caption{\label{II}{Detailed summary of WH properties for Lorentzian noncommutative distribution $\alpha =0.5$, $C_{2}=0.2$, $\theta =0.9$, $ M=0.5$, and $\psi =3.036\times10^{-34}$}}
\begin{tabular}{|c|c|c|c|c|c|c|c|c|}
    \hline
\multicolumn{4}{|c|}{Lorentzian Noncommutative Distribution}\\
    \hline
$Parameter/Expressions$    & $\beta=0.70$          &$\beta=0.90$  &$\beta=1.10$\\
\hline
$\epsilon _s(r)$            & $\epsilon _s(r)>0$ in $0.1\leq r\leq20$   &$\epsilon _s(r)>0$ in $0.1\leq r\leq20$  &$\epsilon _s(r)>0$ in $0.1\leq r\leq20$\\

$\frac{d\epsilon _s}{dr}$   & $\frac{d\epsilon _s}{dr}\mid_{r_{0}}<1$ in $0.1\leq r\leq20$    & $\frac{d\epsilon _s}{dr}\mid_{r_{0}}<1$ in $0.1\leq r\leq20$  &$\frac{d\epsilon _s}{dr}\mid_{r_{0}}<1$ in $0.1\leq r\leq20$\\

$\frac{\epsilon _s}{r}$          & $\frac{\epsilon _s}{r}\rightarrow 0$ as $r\rightarrow\infty$  &$\frac{\epsilon _s}{r}\rightarrow 0$ as $r\rightarrow\infty$  &$\frac{\epsilon _s}{r}\rightarrow 0$ as $r\rightarrow\infty$\\

$\epsilon _{s}-r$                & $r_{0}=0.180$  &$r_{0}=0.165$  &$ r_{0}=0.145$\\
\hline
\end{tabular}
\end{table}
\end{center}

Further, the ECs for the Gaussian distribution are calculated as

\begin{eqnarray}
\rho+p_{r} &=&\frac{1}{8 \pi ^{3/2} (\beta -4)^2 (\beta -3) (\beta +2) \theta ^{5/2} r^4}\times\bigg(16 \pi ^{3/2} \alpha  (\beta -3) (\beta -2) C_{1} \theta ^{5/2} ((\beta -4) (-r))^{-\frac{4}{\beta -4}}\nonumber\\&+&\frac{1}{2} (\beta -4)(\beta -2) (\beta -1) (\beta +2) M r^6 E_{-2-\frac{2}{\beta -4}}\left(\frac{r^2}{4 \theta }\right)+2 (\beta -4)^2 (\beta +2) \theta  r^4 \left(2 \pi ^{3/2} \theta ^{3/2} \psi \right. \nonumber\\&+&\left.(\beta -1) M e^{-\frac{r^2}{4 \theta }}\right)\bigg),\label{19}\\
\rho-p_{r}&=& \frac{\beta -1}{16 \pi ^{3/2} (\beta -4) (\beta -3) \left(\beta ^2+\beta -2\right) \theta ^{5/2} r^3}\times\bigg(8 \pi ^{3/2} \theta ^{5/2} \left(4 \alpha  (\beta -3) (\beta -2) C_{1} ((\beta -4)\right.\nonumber\\&\times&\left.(-r))^{\frac{\beta }{4-\beta }}-(\beta -4) (\beta +2) r^3 \psi \right)-(\beta -2) (\beta -1) (\beta +2) M r^5 E_{-2-\frac{2}{\beta -4}}\left(\frac{r^2}{4 \theta }\right)-8 (\beta -4)\nonumber\\&\times& (\beta +2)
 \theta  M r^3 e^{-\frac{r^2}{4 \theta }}\bigg),\label{20}
\end{eqnarray}

\begin{eqnarray}
\rho+p_{t} &=&\frac{1}{16 \pi ^{3/2} (\beta -4)^2 (\beta -3) (\beta +2) \theta ^{5/2} r^4}\times\bigg(32 \pi ^{3/2} \alpha  (\beta -3) C_{1} \theta ^{5/2} ((\beta -4) (-r))^{-\frac{4}{\beta -4}}+(\beta -4) \nonumber\\&\times&(\beta -1)(\beta +2) M r^6 E_{-2-\frac{2}{\beta -4}}\left(\frac{r^2}{4 \theta }\right)+4 (\beta -4)^2 (\beta +2) \theta  r^4 \left(2 \pi ^{3/2} \theta ^{3/2} \psi +(\beta -1) M e^{-\frac{r^2}{4 \theta }}\right)\bigg),\label{21}\\
\rho-p_{t} &=&\frac{1}{16 \pi ^{3/2} (\beta -4) (\beta -3) (\beta +2) \theta ^{5/2} r^3}\times\bigg(8 \pi ^{3/2} \theta ^{5/2} \left(4 \alpha  (\beta -3) C_{1} ((\beta -4) (-r))^{\frac{\beta }{4-\beta }}-(\beta -4) \right. \nonumber\\&\times&\left.(\beta +2) r^3 \psi \right)-\left(\beta ^2+\beta -2\right) M r^5 E_{-2-\frac{2}{\beta -4}}\left(\frac{r^2}{4 \theta }\right)-8 (\beta -4) (\beta +2) \theta  M r^3 e^{-\frac{r^2}{4 \theta }}\bigg),\label{22}
\end{eqnarray}
\begin{eqnarray}
\rho+p_{r}+2p_{t} &=&\frac{-1}{16 \pi ^{3/2} (\beta -4) (\beta -3) (\beta +2) \theta ^{5/2} r^3}\times\bigg(8 \pi ^{3/2} \theta ^{5/2} \left(4 \alpha  (\beta -3) \beta  C_{1} ((\beta -4) (-r))^{\frac{\beta }{4-\beta }}-3 \right. \nonumber\\&\times&\left.(\beta -4) (\beta +2) r^3 \psi \right)-\beta  \left(\beta ^2+\beta -2\right) M r^5 E_{-2-\frac{2}{\beta -4}}\left(\frac{r^2}{4 \theta }\right)-8 (\beta -4) \beta  (\beta +2) \theta  M r^3 e^{-\frac{r^2}{4 \theta }}\bigg),\label{23}
\end{eqnarray}

On solving Eq. (\ref{14}), we can get the shape function of WH for the Lorentzian source as:
\begin{eqnarray}
\epsilon _s(r)&=&\frac{2 \left(\beta ^2+\beta -2\right) \sqrt{\theta } M \left(-\frac{r^2}{\theta }\right)^{-\frac{2}{\beta -4}} \left(Beta\left(-\frac{r^2}{\theta },\frac{2}{\beta -4}+1,-1\right)-Beta\left(-\frac{r^2}{\theta },\frac{2}{\beta -4}+1,0\right)\right)}{\pi ^2 \alpha  (\beta -4) r}\nonumber\\&+&\frac{(\beta +2) r^3 \psi }{4 \alpha  (\beta -3)}+C_{2} ((\beta -4) (-r))^{\frac{\beta }{4-\beta }}\label{24}.
\end{eqnarray}
where $``Beta"$ is a special function, and it is calculated as $B_{-\frac{r^2}{\theta }}\left(\frac{2}{\beta -4}+1,-1\right)$ and $C_{2}$ is a constant of integration. Now, by using Eq. (\ref{24}) in Eqs. (\ref{9}-\ref{11}), we get energy density and the stress components of Lorentzian source as:
\begin{eqnarray}
\rho &=&\frac{\sqrt{\theta } M}{\pi ^2 \left(\theta +r^2\right)^2},\label{25}\\
p_r&=&\frac{1}{2}\times\bigg(\frac{8 (\beta -2) (\beta -1) \sqrt{\theta } M \left(-\frac{r^2}{\theta }\right)^{-\frac{2}{\beta -4}} \left(B_{-\frac{r^2}{\theta }}\left(1+\frac{2}{\beta -4},0\right)-B_{-\frac{r^2}{\theta }}\left(1+\frac{2}{\beta -4},-1\right)\right)}{\pi ^2 (\beta -4)^2 r^4}\nonumber\\ &+&\frac{\psi }{\beta -3}+\frac{4 \alpha  (\beta -2) C_{2} ((\beta -4) (-r))^{-\frac{4}{\beta -4}}}{(\beta -4)^2 (\beta +2) r^4}+\frac{6 \beta  \sqrt{\theta } M}{\pi ^2 (\beta -4) \left(\theta +r^2\right)^2}\bigg),\label{26}\\
p_t&=&\frac{4 (\beta -1) \sqrt{\theta } M \left(-\frac{r^2}{\theta }\right)^{-\frac{2}{\beta -4}} \left(B_{-\frac{r^2}{\theta }}\left(1+\frac{2}{\beta -4},0\right)-B_{-\frac{r^2}{\theta }}\left(1+\frac{2}{\beta -4},-1\right)\right)}{\pi ^2 (\beta -4)^2 r^4}\nonumber\\ &+&\frac{\psi }{2 (\beta -3)}+\frac{2 \alpha  C_{2} ((\beta -4) (-r))^{-\frac{4}{\beta -4}-2}}{(\beta +2) r^2}+\frac{(\beta +2) \sqrt{\theta } M}{\pi ^2 (\beta -4) \left(\theta +r^2\right)^2}.\label{27}
\end{eqnarray}

Further, the ECs for the Lorentzian distribution are calculated as

\begin{eqnarray}
\rho+p_{r} &=&\frac{-1}{2 \pi ^2 (\beta -4)^2 r^4 \left(\theta +r^2\right)^2}\times\bigg(\frac{((\beta -4) (-r))^{-\frac{4}{\beta -4}}}{(\beta -3) (\beta +2)}\bigg((\beta +2) r^3 ((\beta -4) (-r))^{\frac{\beta }{\beta -4}} \left(8 (\beta -3)\right.\nonumber\\&\times&\left. (\beta -1) \sqrt{\theta } M+\pi ^2 (\beta -4) \psi  \left(\theta +r^2\right)^2\right)-4 \pi ^2 \alpha  (\beta -3) (\beta -2) C_{2} \left(\theta +r^2\right)^2\bigg)+8 (\beta -2)\nonumber\\&\times& (\beta -1) M\sqrt{\theta } \left(-\frac{r^2}{\theta }\right)^{-\frac{6}{\beta -4}}\bigg(\left(\theta +r^2\right)^2 \left(-\frac{r^2}{\theta }\right)^{\frac{4}{\beta -4}} B_{-\frac{r^2}{\theta }}\left(1+\frac{2}{\beta -4},-1\right)-\left(\left(\theta ^2+r^4\right) \right.\nonumber\\&\times&\left.\left(-\frac{r^2}{\theta }\right)^{\frac{4}{\beta -4}}-2 \theta ^2 \left(-\frac{r^2}{\theta }\right)^{\frac{\beta }{\beta -4}}\right) B_{-\frac{r^2}{\theta }}\left(1+\frac{2}{\beta -4},0\right)\big)\bigg),\label{28}\\
\rho-p_{r}&=&\frac{1}{2 \pi ^2 (\beta -4) r^3} \times\bigg(\frac{1}{(\beta -3) (\beta +2) \left(\theta +r^2\right)^2}\bigg(4 \pi ^2 \alpha  (\beta -3) (\beta -2) C_{2} ((\beta -4) (-r))^{-\frac{\beta }{\beta -4}}\nonumber\\&+& \left(\theta +r^2\right)^2+(\beta +2) r^3 \left(-4 (\beta -3) (\beta +2) \sqrt{\theta } M-\pi ^2 (\beta -4) \psi  \left(\theta +r^2\right)^2\right)\bigg)+\frac{8 (\beta -1) M r}{\sqrt{\theta }}\nonumber\\&\times&\bigg(\, _2F_1\left(1,\frac{2}{\beta -4}+1;\frac{2}{\beta -4}+2;-\frac{r^2}{\theta }\right)-\, _2F_1\left(2,1+\frac{2}{\beta -4};2+\frac{2}{\beta -4};-\frac{r^2}{\theta }\right)\bigg)\bigg),\label{29}
\end{eqnarray}
\begin{eqnarray}
\rho+p_{t} &=&\frac{1}{2 \pi ^2 (\beta -4)^2 r^4}\times\bigg(\frac{4 \pi ^2 \alpha  C_{2} ((\beta -4) (-r))^{-\frac{4}{\beta -4}}}{\beta +2}-8 (\beta -1) \sqrt{\theta } M \left(-\frac{r^2}{\theta }\right)^{-\frac{2}{\beta -4}}\nonumber\\&\times& B_{-\frac{r^2}{\theta }}\left(1+\frac{2}{\beta -4},-1\right)+(\beta -4) r^4 \left(\frac{\pi ^2 (\beta -4) \psi }{\beta -3}+\frac{4 (\beta -1) \sqrt{\theta } M}{\left(\theta +r^2\right)^2}\right)+\frac{1}{(\beta -3) (\beta +2)}\nonumber\\&\times&\frac{8 \sqrt{\theta } M \left(-\frac{r^2}{\theta }\right)^{-\frac{6}{\beta -4}}}{\left(\theta +r^2\right)^2}\times B_{-\frac{r^2}{\theta }}\left(\frac{2}{\beta -4}+1,0\right)\bigg(2 \beta  (2 \beta +5) \theta ^2 \left(-\frac{r^2}{\theta }\right)^{\frac{\beta }{\beta -4}}+\left((\beta -3) \right.\nonumber\\&\times&\left.(\beta -1) (\beta +2) \theta ^2+(\beta -3) (\beta -1) (\beta +2) r^4+2 \left(\beta ^3+6\right) \theta  r^2\right) \left(-\frac{r^2}{\theta }\right)^{\frac{4}{\beta -4}}\bigg)\bigg),\label{30}\\
\rho-p_{t} &=&\frac{1}{2 \pi ^2 (\beta -4)^2 r^4}\times\bigg(\frac{-(\beta -4) r}{(\beta -3) (\beta +2) \left(\theta +r^2\right)^2}\bigg((\beta +2) r^3 \left(12 (\beta -3) \sqrt{\theta } M+\pi ^2 (\beta -4) \psi\right. \nonumber\\&\times&\left. \left(\theta +r^2\right)^2\right)-4 \pi ^2 \alpha  (\beta -3) C_{2} ((\beta -4) (-r))^{-\frac{\beta }{\beta -4}} \left(\theta +r^2\right)^2\bigg)+8 (\beta -1) \sqrt{\theta } M \left(-\frac{r^2}{\theta }\right)^{-\frac{2}{\beta -4}}\nonumber\\&\times&\bigg(B_{-\frac{r^2}{\theta }}\left(1+\frac{2}{\beta -4},-1\right)-B_{-\frac{r^2}{\theta }}\left(1+\frac{2}{\beta -4},0\right)\bigg)\bigg),\label{31}\\
\rho+p_{r}+2p_{t} &=&\frac{-1}{2 \pi ^2(\beta -4)^2 r^4}\times\bigg(3(\beta -4) r^4\left(-\frac{\pi ^2(\beta -4) \psi }{\beta -3}-\frac{4 \beta  \sqrt{\theta } M}{\left(\theta +r^2\right)^2}\right)-\frac{1}{\beta +2}\bigg(4 \pi ^2 \alpha  \beta  C_{2} ((\beta -4) \nonumber\\&\times&(-r))^{-\frac{4}{\beta -4}}\bigg)+8(\beta -1)\beta \sqrt{\theta } M \left(-\frac{r^2}{\theta }\right)^{-\frac{2}{\beta -4}}\left(B_{-\frac{r^2}{\theta }}\left(1+\frac{2}{\beta -4},-1\right)\right.\nonumber\\&-&\left.B_{-\frac{r^2}{\theta }}\left(1+\frac{2}{\beta -4},0\right)\right)\bigg),\label{32}
\end{eqnarray}
where $\, _2F_1$ is hypergeometric special function.

\begin{figure}
\centering \epsfig{file=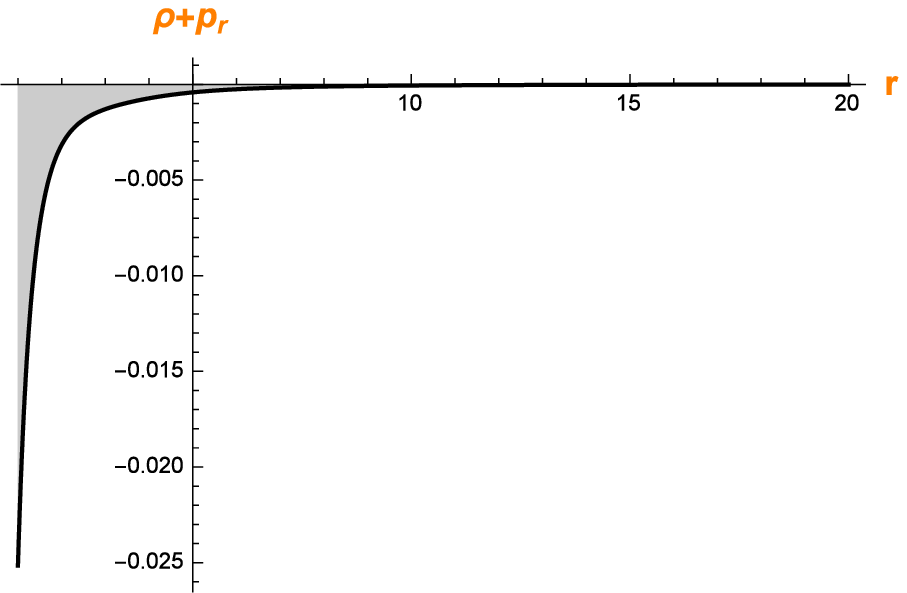, width=.32\linewidth,
height=2.02in}\epsfig{file=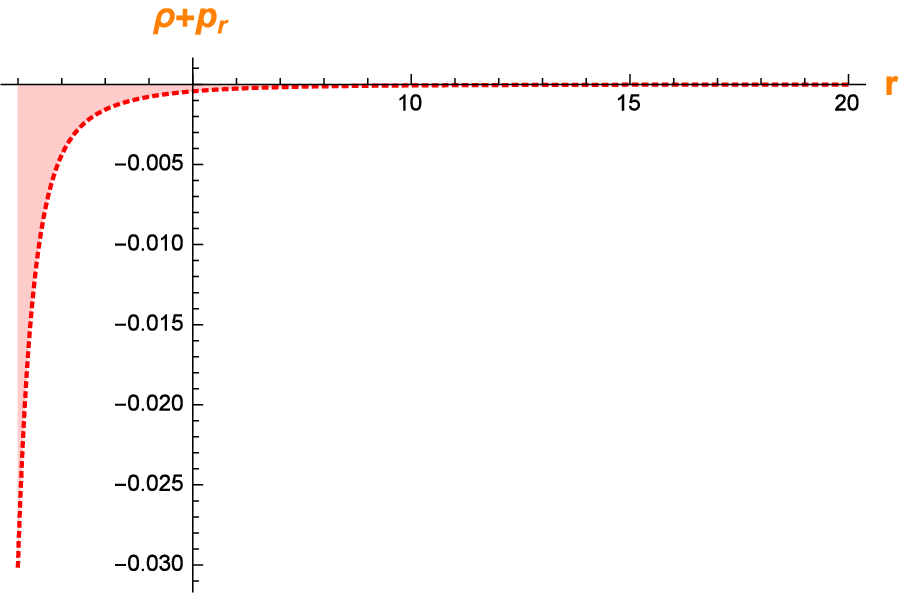, width=.32\linewidth,
height=2.02in}\epsfig{file=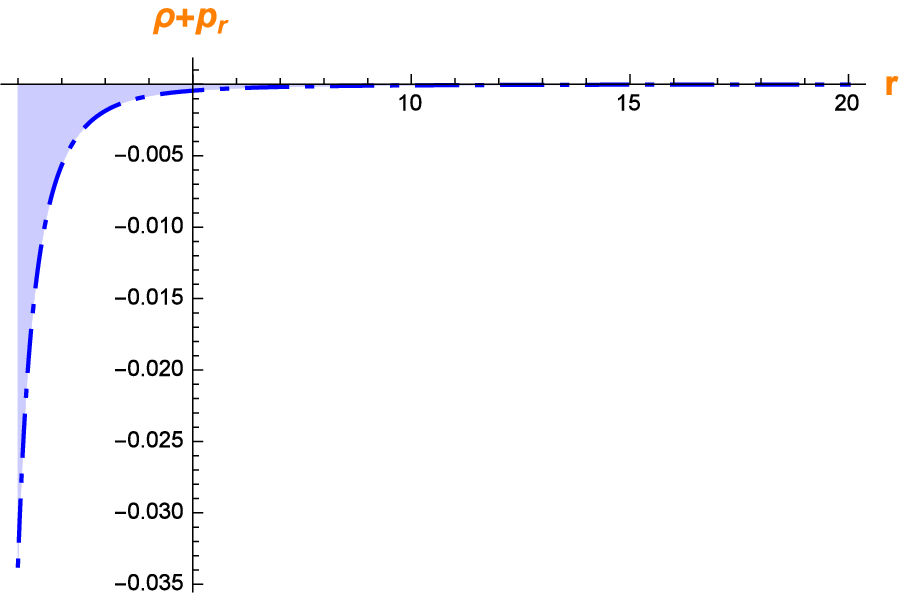, width=.32\linewidth,
height=2.02in}
\centering \epsfig{file=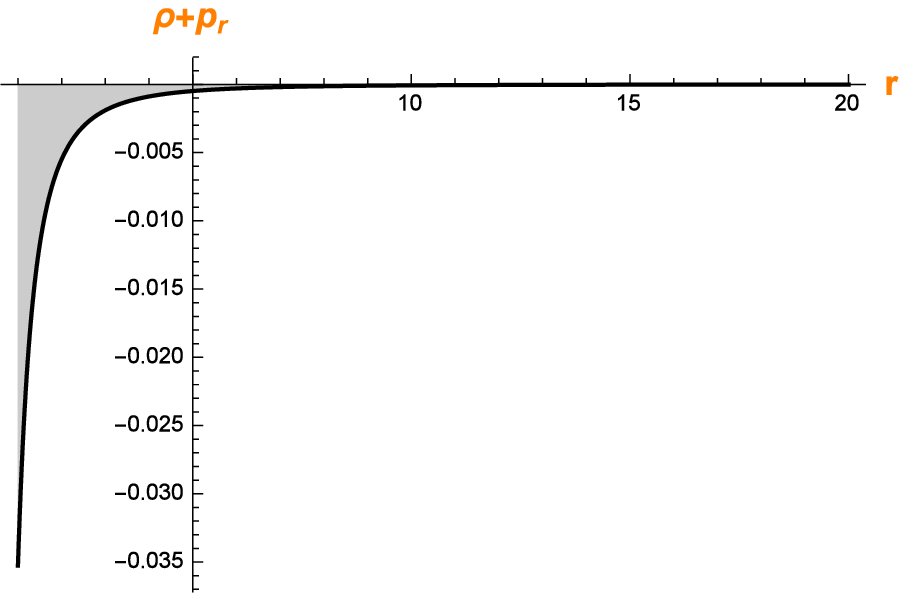, width=.32\linewidth,
height=2.02in}\epsfig{file=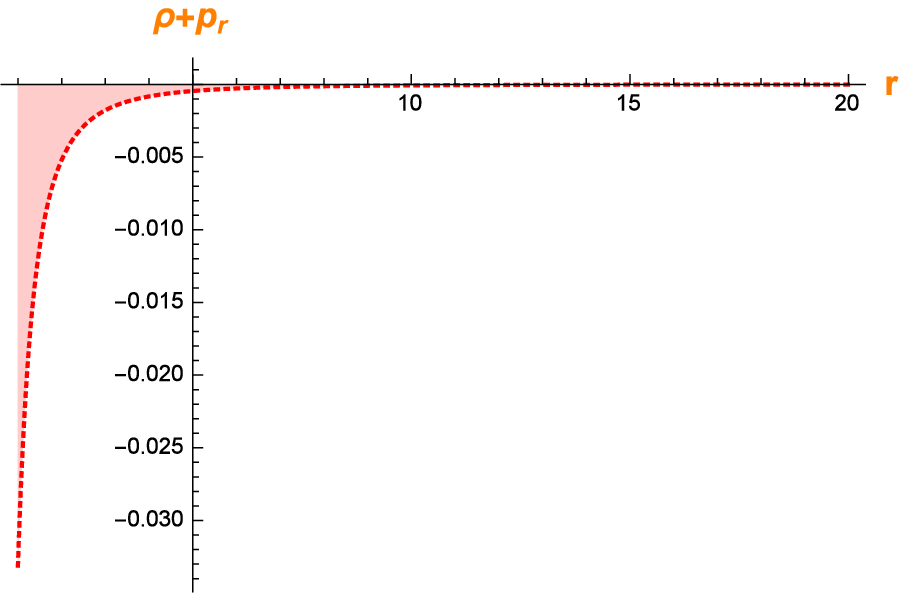, width=.32\linewidth,
height=2.02in}\epsfig{file=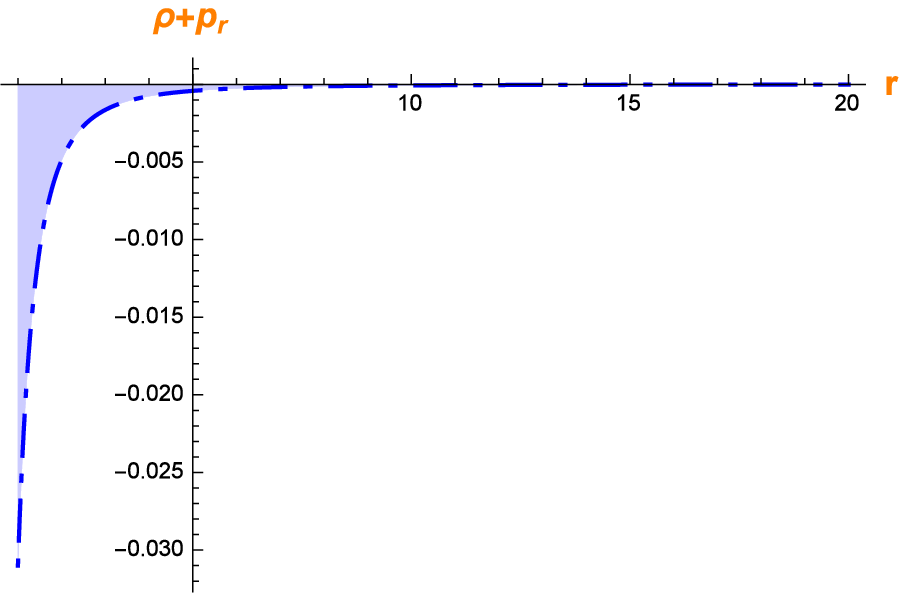, width=.32\linewidth,
height=2.02in} \caption{\label{fig5} Shows the behavior of $\rho+p_{r}$.}
\end{figure}

\begin{figure}
\centering \epsfig{file=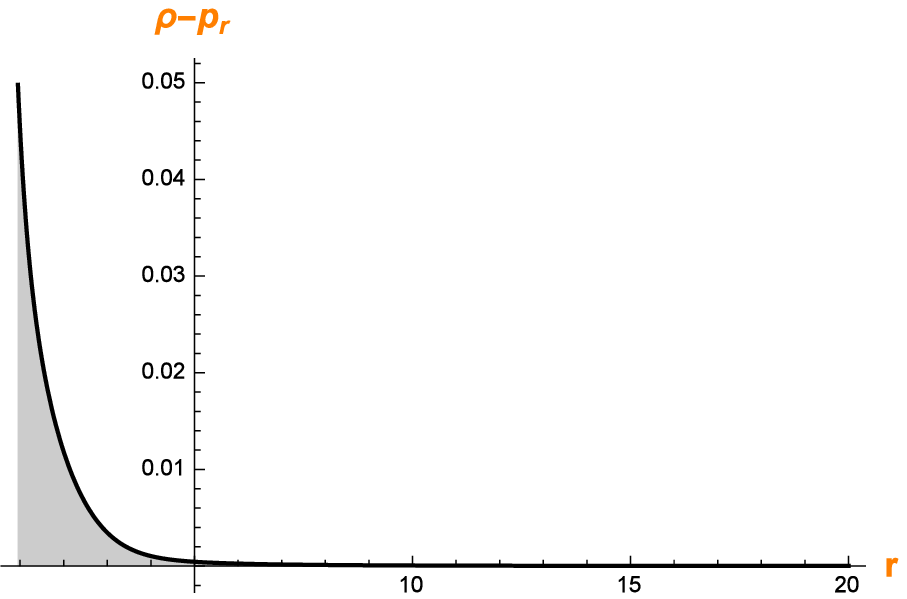, width=.32\linewidth,
height=2.02in}\epsfig{file=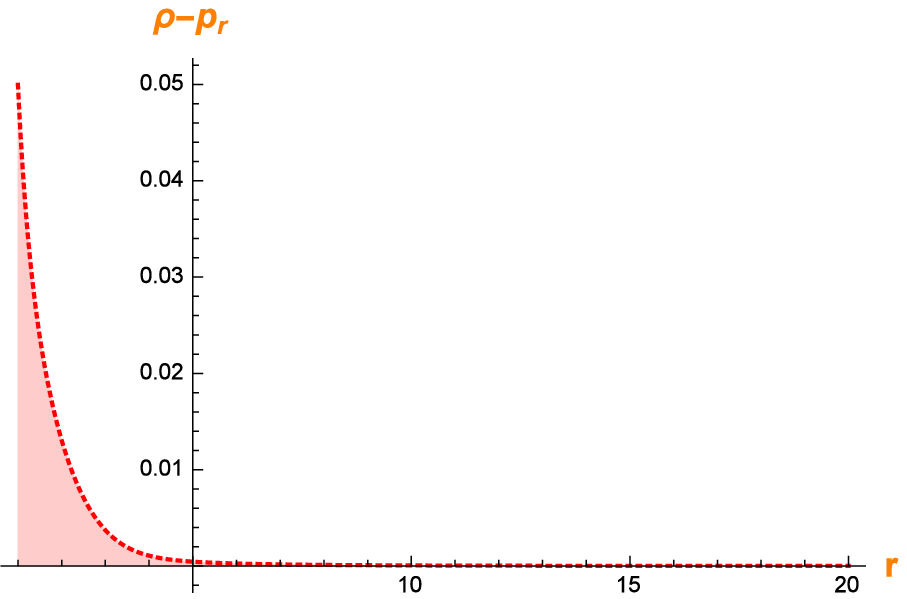, width=.32\linewidth,
height=2.02in}\epsfig{file=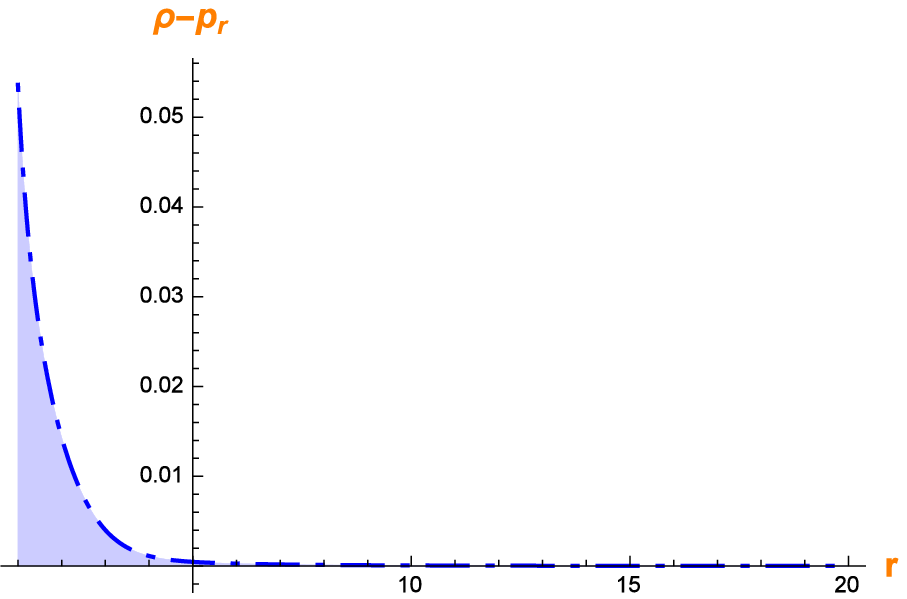, width=.32\linewidth,
height=2.02in}
\centering \epsfig{file=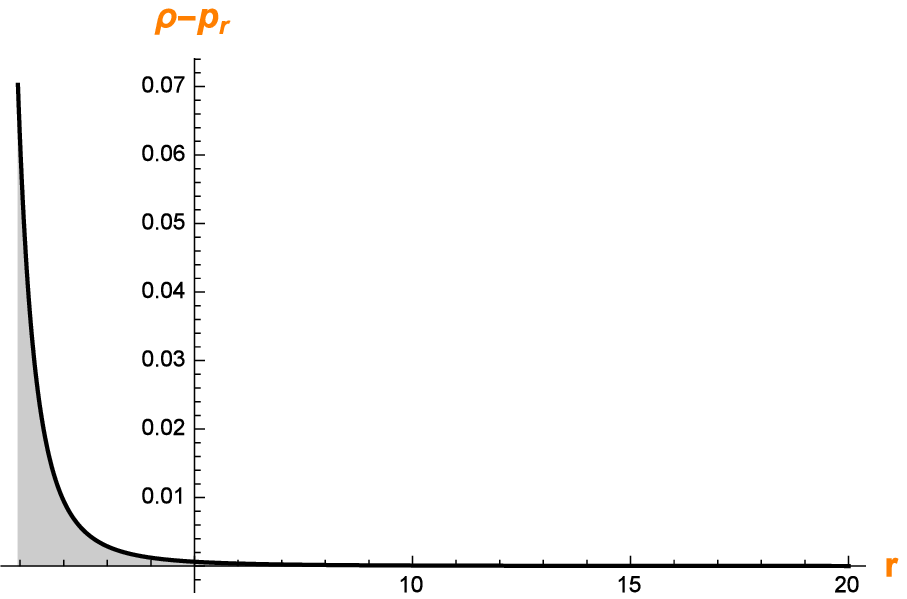, width=.32\linewidth,
height=2.02in}\epsfig{file=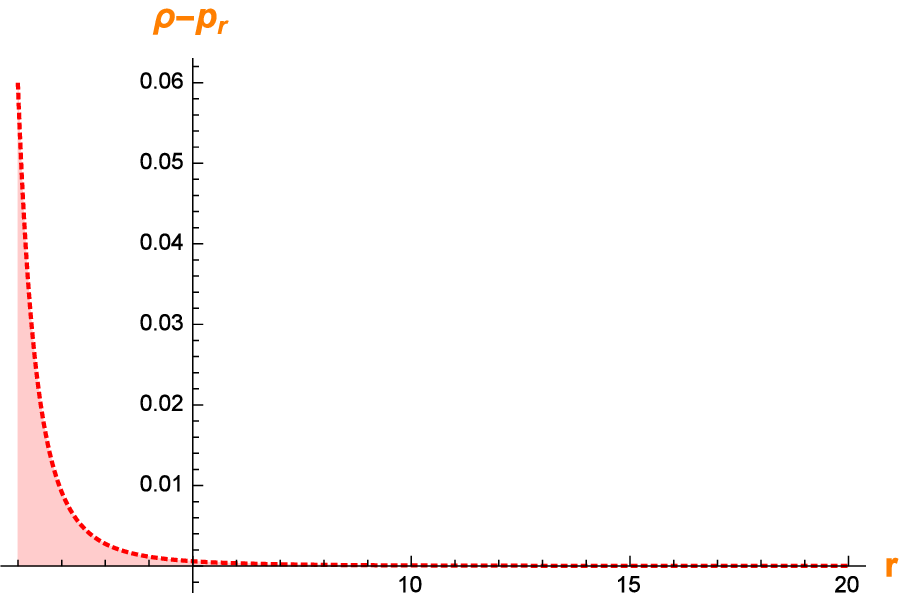, width=.32\linewidth,
height=2.02in}\epsfig{file=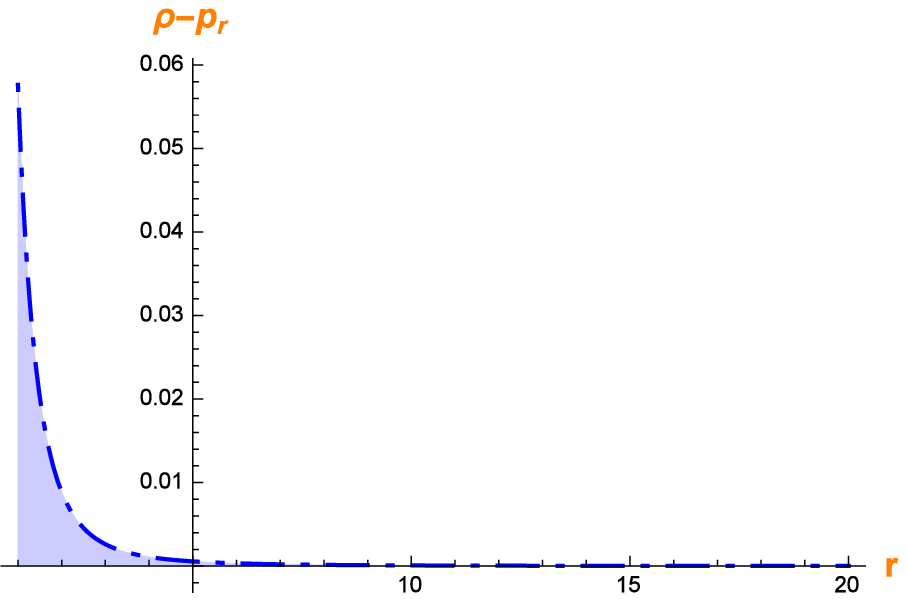, width=.32\linewidth,
height=2.02in} \caption{\label{fig6} Shows the behavior of $\rho-p_{r}$.}
\end{figure}

\begin{figure}
\centering \epsfig{file=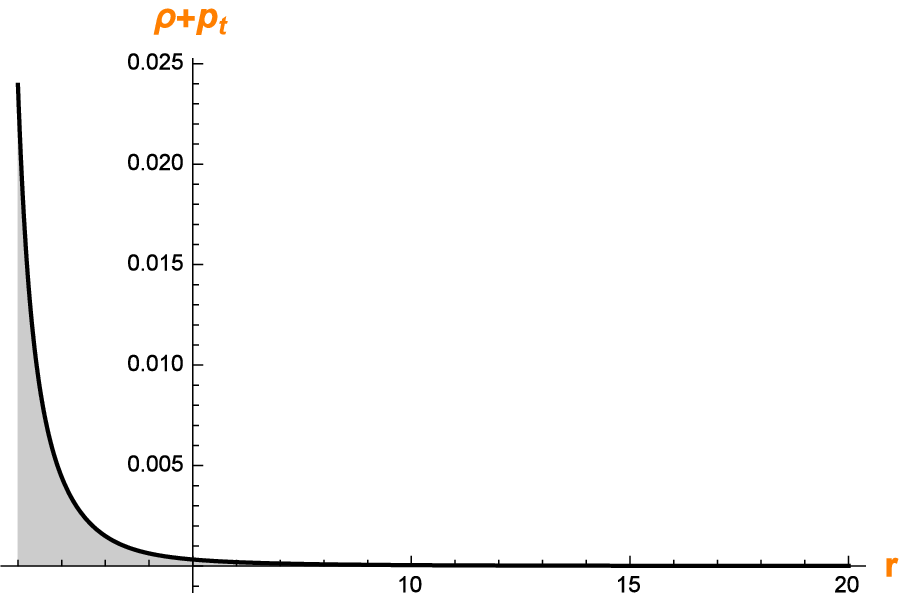, width=.32\linewidth,
height=2.02in}\epsfig{file=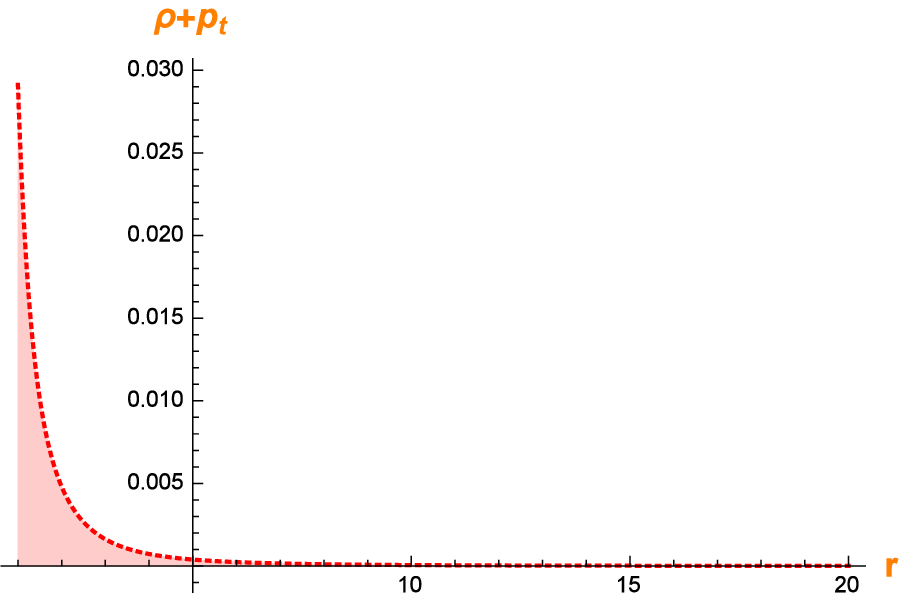, width=.32\linewidth,
height=2.02in}\epsfig{file=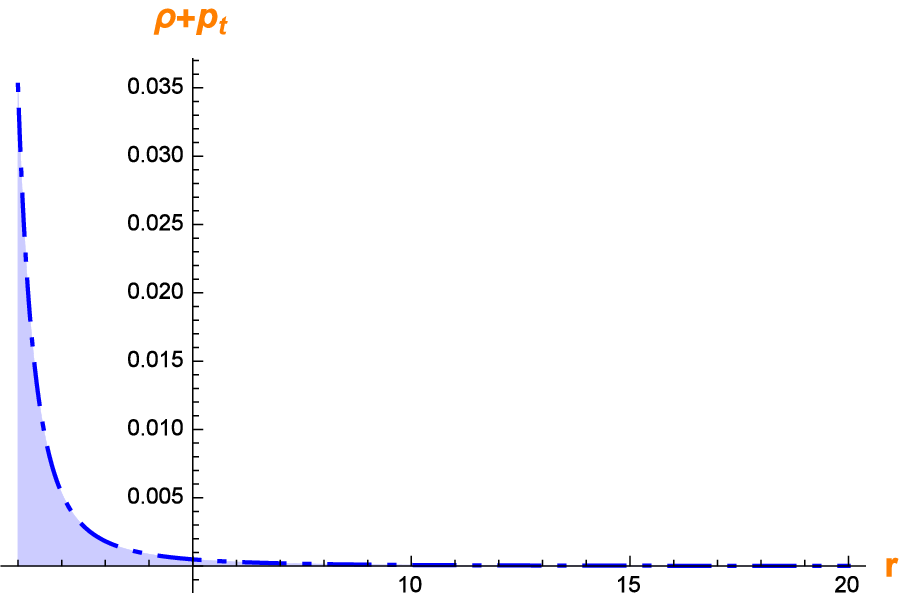, width=.32\linewidth,
height=2.02in}
\centering \epsfig{file=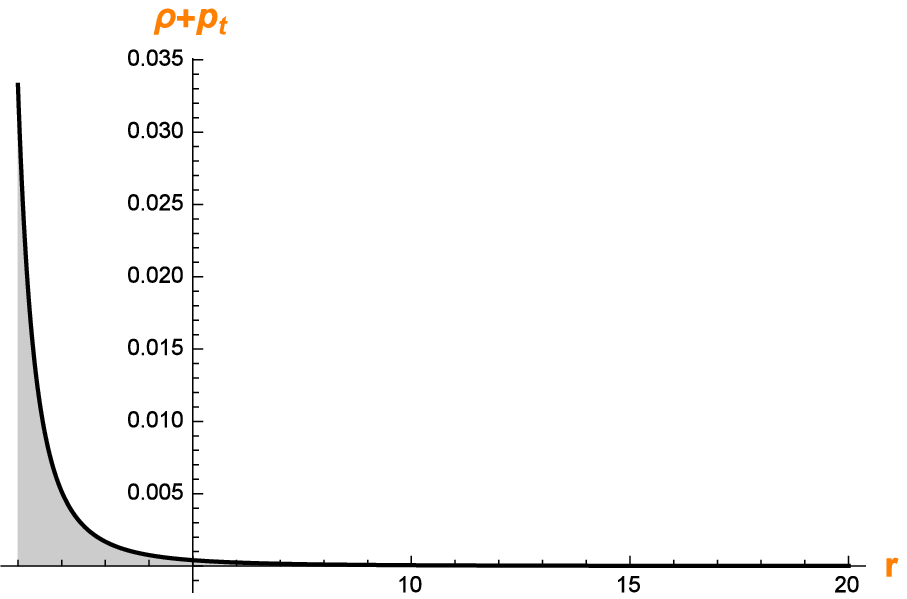, width=.32\linewidth,
height=2.02in}\epsfig{file=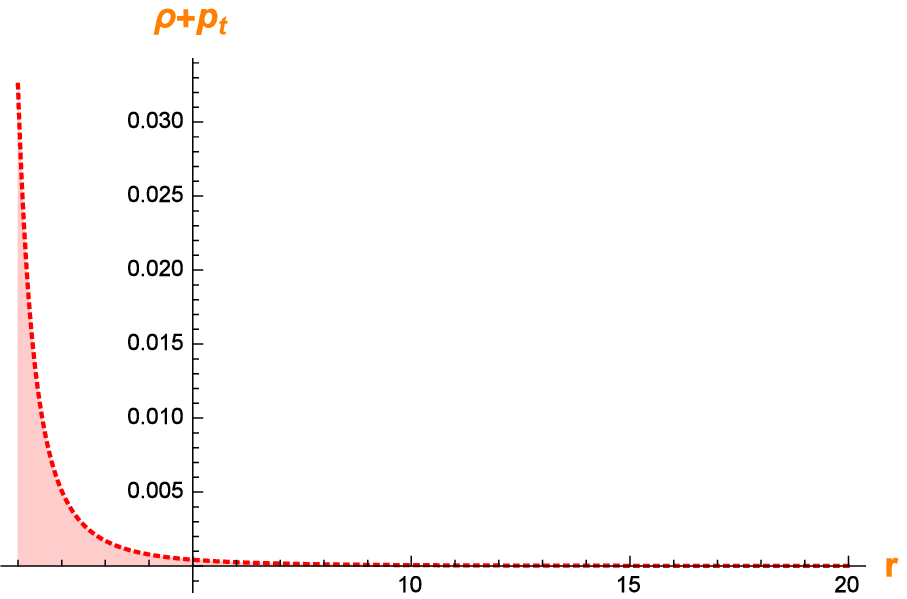, width=.32\linewidth,
height=2.02in}\epsfig{file=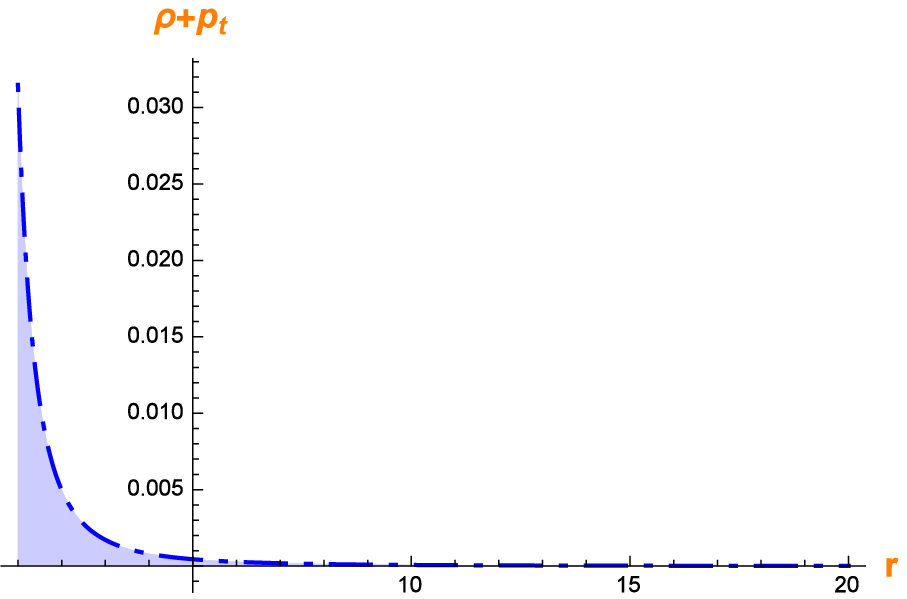, width=.32\linewidth,
height=2.02in} \caption{\label{fig7} Shows the behavior of $\rho+p_{t}$.}
\end{figure}

\begin{figure}
\centering \epsfig{file=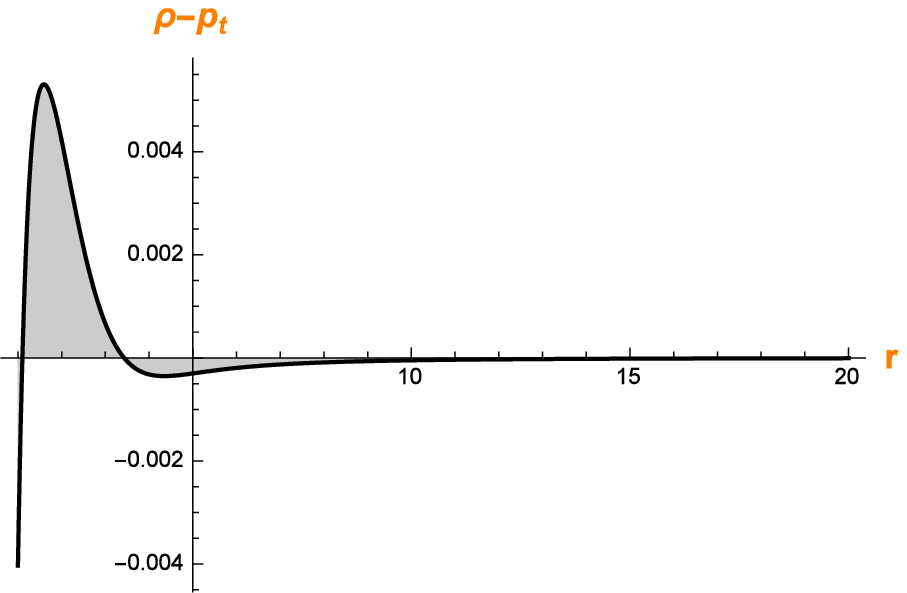, width=.32\linewidth,
height=2.02in}\epsfig{file=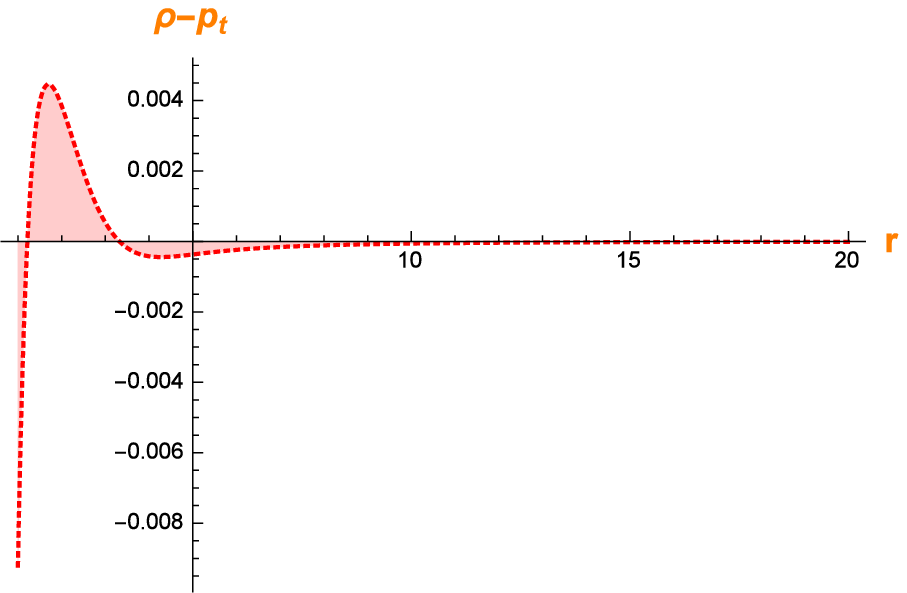, width=.32\linewidth,
height=2.02in}\epsfig{file=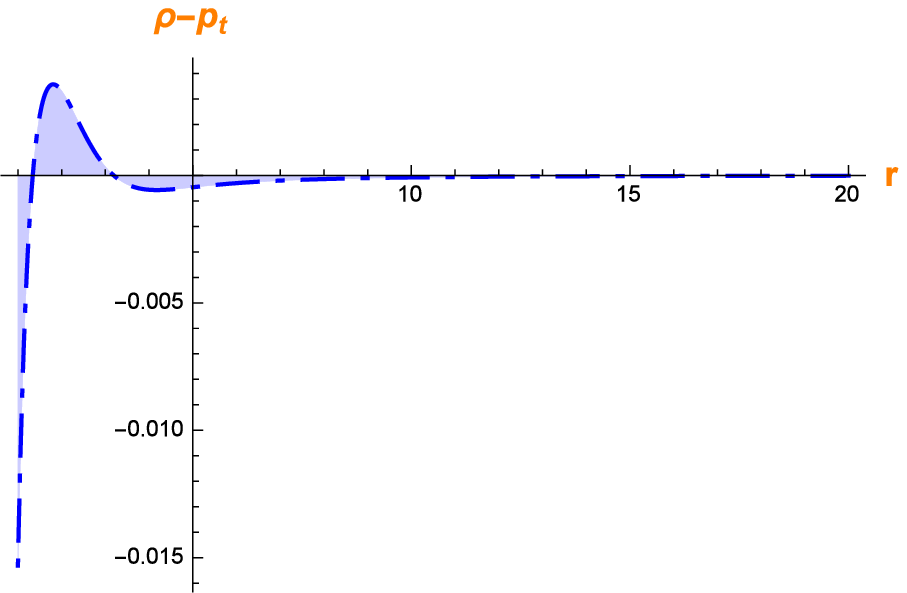, width=.32\linewidth,
height=2.02in}
\centering \epsfig{file=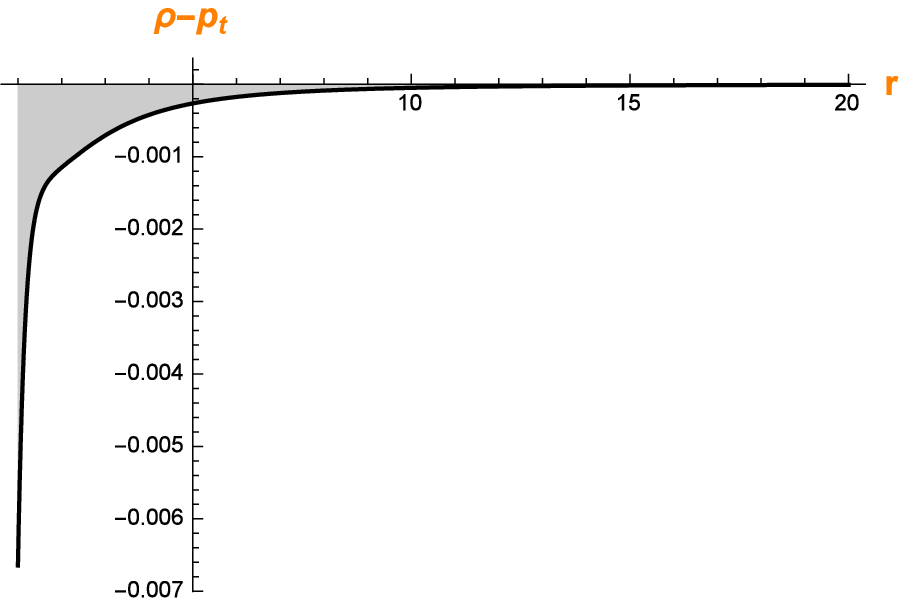, width=.32\linewidth,
height=2.02in}\epsfig{file=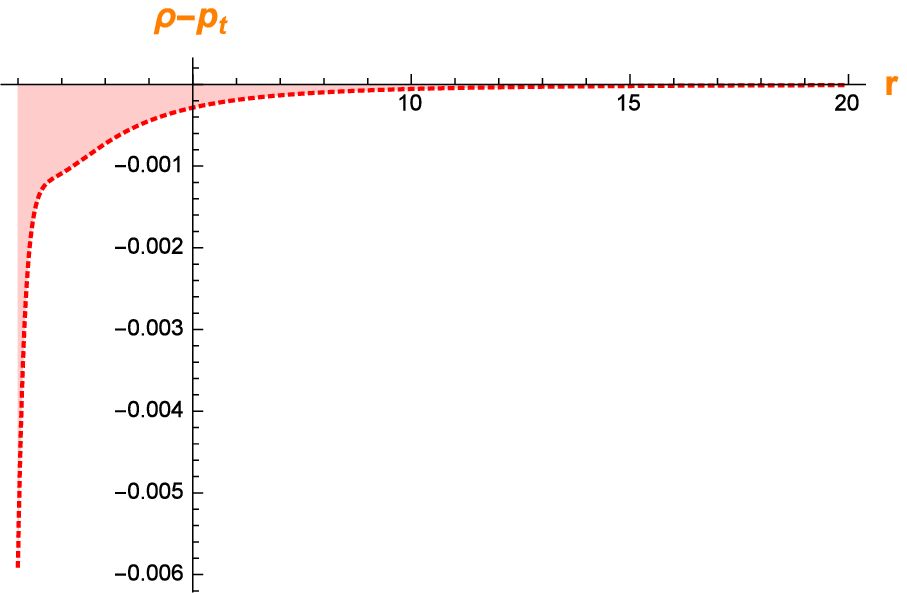, width=.32\linewidth,
height=2.02in}\epsfig{file=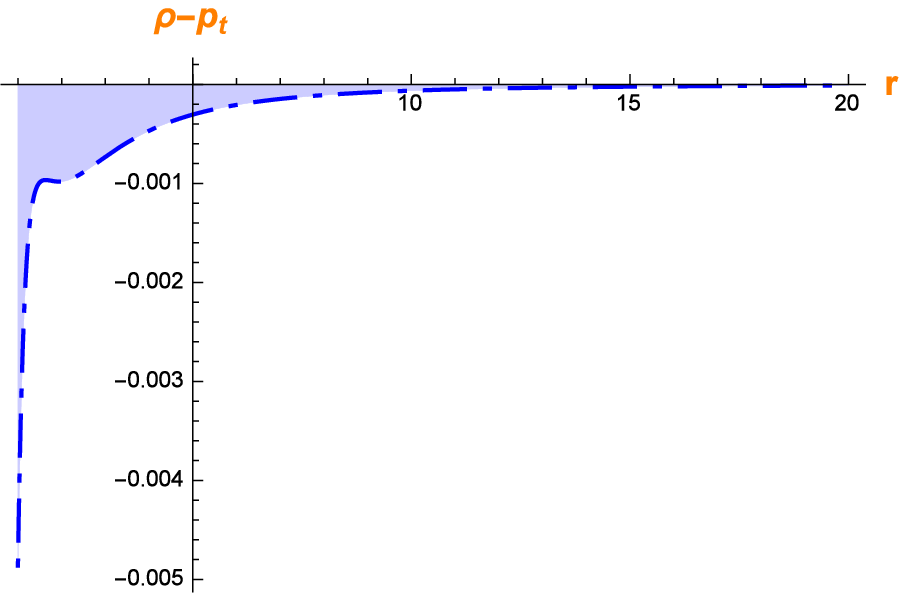, width=.32\linewidth,
height=2.02in} \caption{\label{fig8} Shows the behavior of $\rho-p_{t}$.}
\end{figure}

\begin{figure}
\centering \epsfig{file=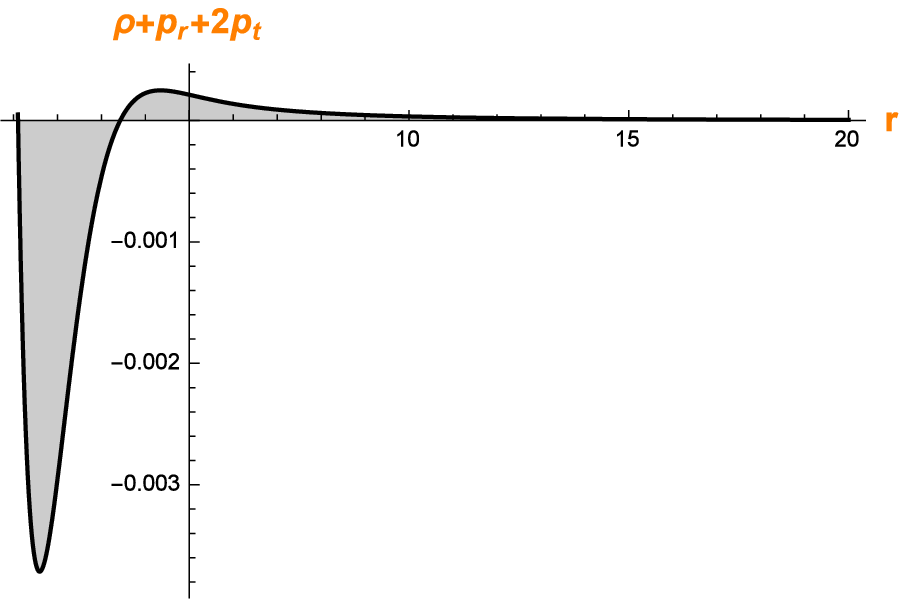, width=.32\linewidth,
height=2.02in}\epsfig{file=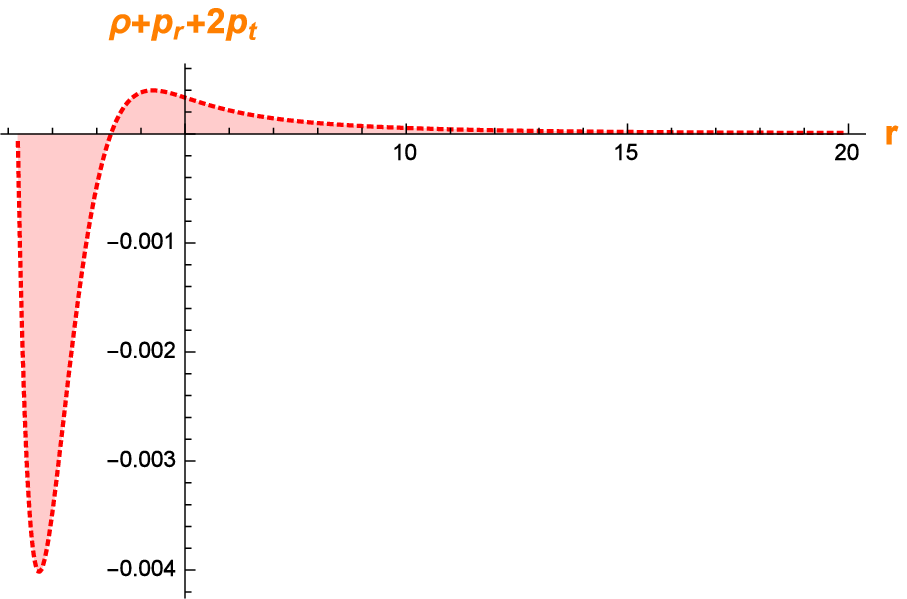, width=.32\linewidth,
height=2.02in}\epsfig{file=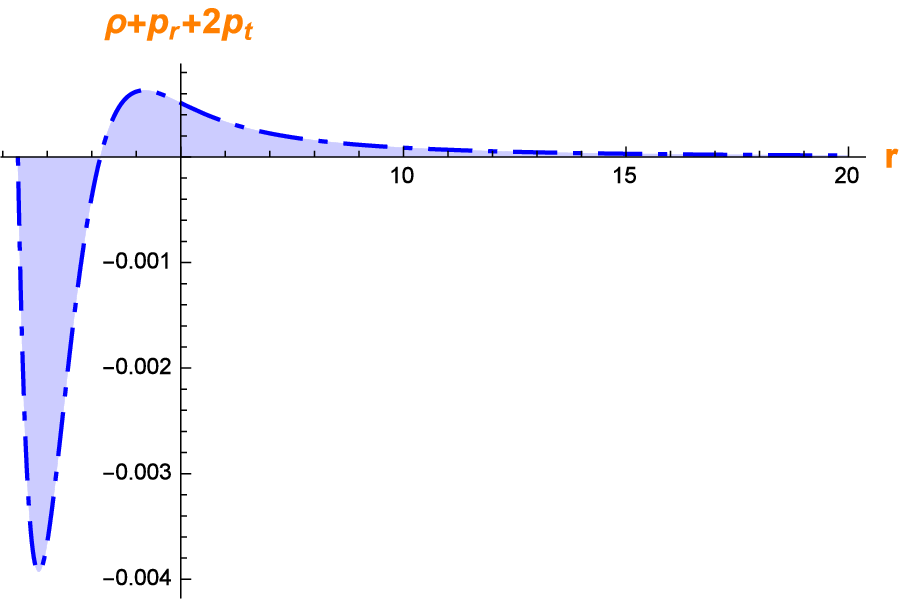, width=.32\linewidth,
height=2.02in}
\centering \epsfig{file=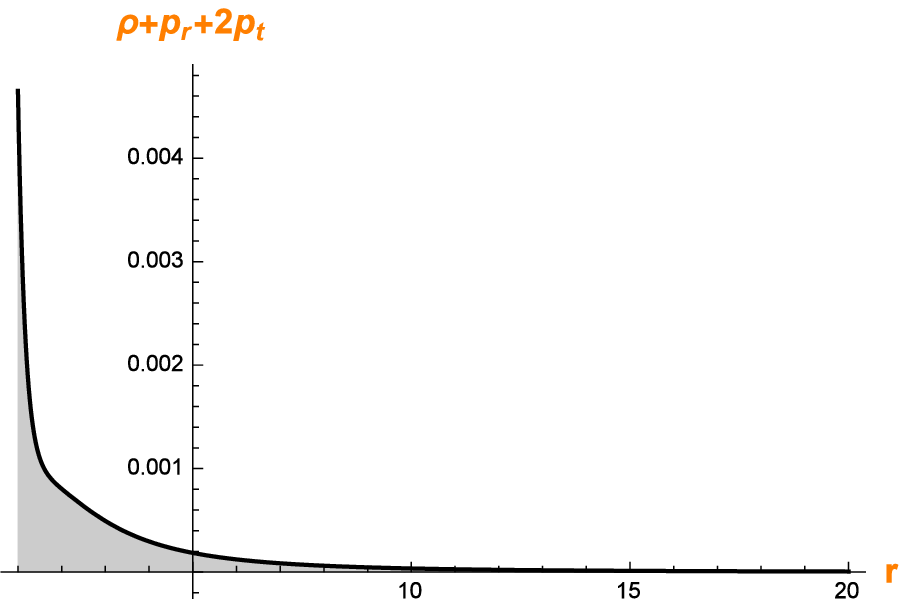, width=.32\linewidth,
height=2.02in}\epsfig{file=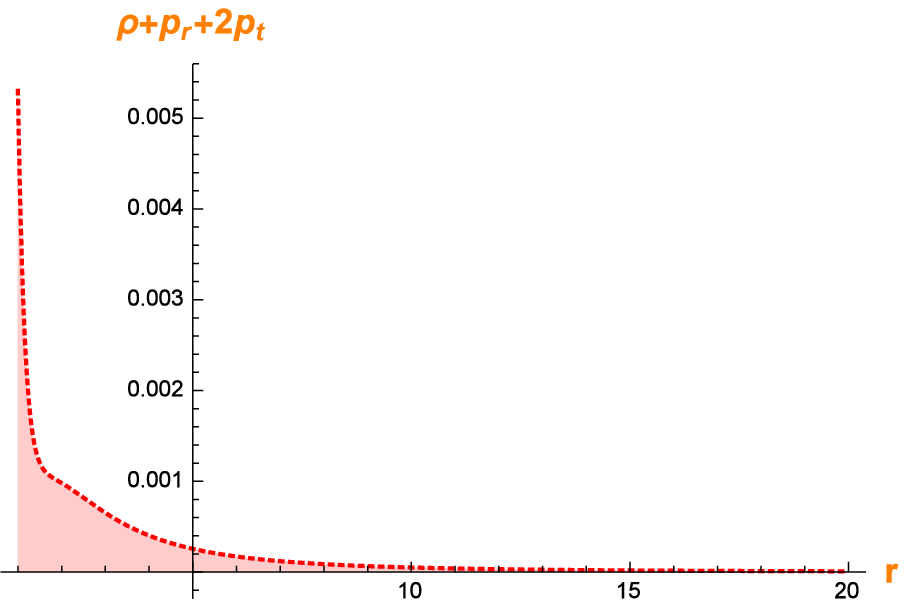, width=.32\linewidth,
height=2.02in}\epsfig{file=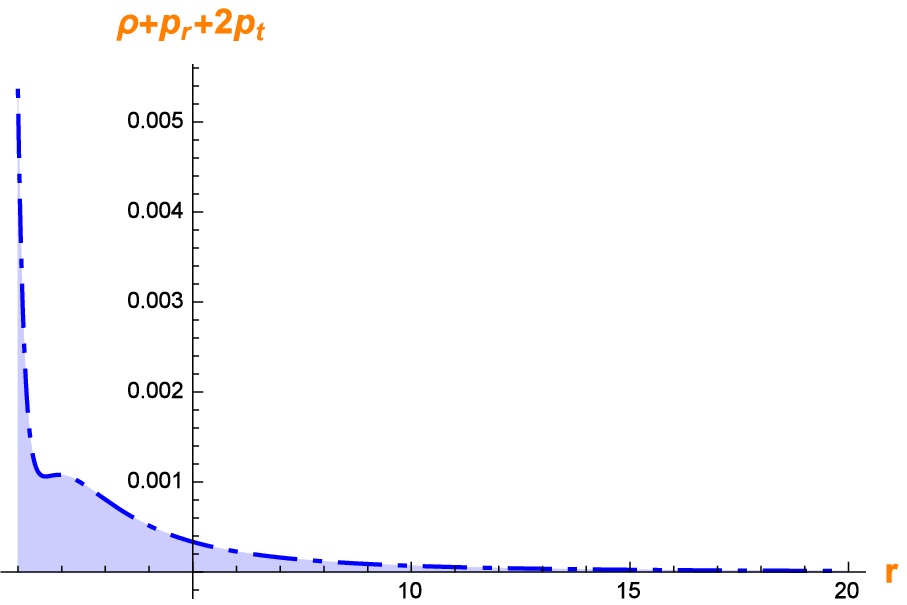, width=.32\linewidth,
height=2.02in} \caption{\label{fig9} Shows the behavior of $\rho+p_{r}+2p_{t}$.}
\end{figure}
The ECs have a critical role in the WH study. In this regard, we shall calculate the expressions for $\rho+p_{r}$, $\rho-p_{r}$, $\rho+p_{t}$, $\rho-p_{t}$, and $\rho+p_{r}+2p_{t}$. The graphical behavior of ECs is provided in Figs. (\ref{5}-\ref{9}) for the Gaussian and the Lorentzian distributions. From the Fig. (\ref{5}), the $NEC$ has been seen violated, i.e., $\rho+p_{r}<0$ within $1\leq r\leq20$ in both the cases for $\beta=0.70,\;0.90,\;\&\;1.10$, which is compatible with the non-commutative geometries. This violation of $(NEC)$ confirms the presence of exotic matter in both the cases which is necessary requirement for the WH existence. The positive graphical behavior of $\rho-p_{r}$ can be noticed from the Fig. (\ref{6}) for both Gaussian (first row) and Lorentzian (second row) under the similar parametric constraints and is reflected in the left, middle, and right portions, respectively. The graphical conduct of $\rho+p_{t}$ for both the Gaussian (first row) and Lorentzian (second row) is described in the Fig. (\ref{7}). It is evident from the Fig. (\ref{7}) that $\rho+p_{t}$ remains positive, i.e., $\rho+p_{t}>0$ within $1\leq r\leq20$. The negative conduct of $\rho-p_{t}$ can be noticed from Fig. (\ref{8}) for both the distributions. Further, the graphical behavior of $\rho+p_{r}+2p_{t}$ expression can be revealed from the Fig. (\ref{9}) for both the Gaussian (first row) and the Lorentzian (second row) under $\beta=0.70,\;0.90,\;\&\;1.10$ by left, middle, and right parts, respectively. All the calculated results of ECs in this scenario are provided in Tabs. (\ref{III}-\ref{IV}).

\begin{center}
\begin{table}
\caption{\label{III}{Detailed summary of ECs for Gaussian non-commutative distribution $\alpha =0.5$, $C_{1}=0.2$, $\theta =0.9$, $ M=0.5$, and $\psi =3.036\times10^{-34}$}}
\begin{tabular}{|c|c|c|c|c|c|c|c|c|}
    \hline
\multicolumn{4}{|c|}{Gaussian Noncommutative Distribution}\\
    \hline
$Parameter/Expressions$    & $\beta=0.70$          &$\beta=0.90$  &$\beta=1.10$\\
\hline
$\rho+p_{r}$               & $\rho+p_{r}<0$ in $1\leq r\leq20$   &$\rho+p_{r}<0$ in $1\leq r\leq20$   &$\rho+p_{r}<0$ in $1\leq r\leq20$\\

$\rho-p_{r}$               & $\rho+p_{r}>0$ in $1\leq r\leq20$   &$\rho+p_{r}>0$ in $1\leq r\leq20$   &$\rho+p_{r}>0$ in $1\leq r\leq20$\\

$\rho+p_{t}$               & $\rho+p_{t}>0$ in $1\leq r\leq20$   &$\rho+p_{t}>0$ in $1\leq r\leq20$   &$\rho+p_{t}>0$ in $1\leq r\leq20$\\

$\rho-p_{t}$               & $\rho-p_{t}<0$ in $3\leq r\leq20$   &$\rho-p_{t}<0$ in $3\leq r\leq20$   &$\rho-p_{t}<0$ in $3\leq r\leq20$\\
$\rho+p_{r}+2p_{t}$        & $\rho+p_{r}+2p_{t}<0$ in $1\leq r\leq3.5$   &$\rho+p_{r}+2p_{t}<0$ in $1\leq r\leq3.3$   &$\rho+p_{r}+2p_{t}<0$ in $1\leq r\leq3.1$ \\
\hline
\end{tabular}
\end{table}
\end{center}
\begin{center}
\begin{table}
\caption{\label{IV}{Detailed summary of ECs for Lorentzian non-commutative distribution $\alpha =0.5$, $C_{2}=0.2$, $\theta =0.9$, $ M=0.5$, and $\psi =3.036\times10^{-34}$}}
\begin{tabular}{|c|c|c|c|c|c|c|c|c|}
    \hline
\multicolumn{4}{|c|}{Lorentzian Noncommutative Distribution}\\
    \hline
$Parameter/Expressions$    & $\beta=0.70$          &$\beta=0.90$  &$\beta=1.10$\\
\hline
$\rho+p_{r}$               & $\rho+p_{r}<0$ in $1\leq r\leq20$   &$\rho+p_{r}<0$ in $1\leq r\leq20$   &$\rho+p_{r}<0$ in $1\leq r\leq20$\\

$\rho-p_{r}$               & $\rho+p_{r}>0$ in $1\leq r\leq20$   &$\rho+p_{r}>0$ in $1\leq r\leq20$   &$\rho+p_{r}>0$ in $1\leq r\leq20$\\

$\rho+p_{t}$               & $\rho+p_{t}>0$ in $1\leq r\leq20$   &$\rho+p_{t}>0$ in $1\leq r\leq20$   &$\rho+p_{t}>0$ in $1\leq r\leq20$\\

$\rho-p_{t}$               & $\rho-p_{t}<0$ in $1\leq r\leq20$   &$\rho-p_{t}<0$ in $1\leq r\leq20$   &$\rho-p_{t}<0$ in $1\leq r\leq20$\\
$\rho+p_{r}+2p_{t}$        & $\rho+p_{r}+2p_{t}>0$ in $1\leq r\leq20$ &$\rho+p_{r}+2p_{t}>0$ in $1\leq r\leq20$  &$\rho+p_{r}+2p_{t}>0$ in $1\leq r\leq20$  \\
\hline
\end{tabular}
\end{table}
\end{center}

\section{Stability of Gaussian and Lorentzian wormholes models}
Here, we test the stability of our obtained solutions by utilizing Tolman-Oppenheimer-Volkov (TOV) equation with one extra term due to the matter coupling, which is described as
\begin{eqnarray}\label{33}
\frac{\varpi^{'}}{2}(\rho+p_r)+\frac{dp_r}{dr}+\frac{2}{r}(p_r-p_t)+\left(\frac{-2 \beta }{\beta +1}\times \bigg(\frac{1}{4}\frac{d\rho}{dr}-\frac{1}{4}\frac{dp_r}{dr}-\frac{dp_t}{dr}\bigg)\right)=0,
\end{eqnarray}
where
\begin{eqnarray}\label{34}
F_h=\frac{dp_r}{dr}, ~~~~~F_g=-\frac{\varpi^{'}}{2}(\rho+p_r), ~~~~F_a=\frac{2}{r}(p_t-p_r), ~~~~~F_e=\left(\frac{-2 \beta }{\beta +1}\times \bigg(\frac{1}{4}\frac{d\rho}{dr}-\frac{1}{4}\frac{dp_r}{dr}-\frac{dp_t}{dr}\bigg)\right)
\end{eqnarray}
In Eq. (\ref{34}), the terms $F_g$, $F_a$, $F_h$ and $F_e$ represent the gravitational, anisotropic, hydrostatic, and extra forces. The force $F_g$ should be zero in both the Gaussian and the Lorentzian distributions. For Gaussian distribution the $F_a$, $F_h$ and $F_e$ forces are calculated as
\begin{eqnarray}
F_a &=&-\frac{4 \alpha  (\beta -3) C_{1} ((\beta -4) (-r))^{-\frac{4}{\beta -4}}}{(\beta -4)^2 (\beta +2) r^5}-\frac{(\beta -1) M r E_{-2-\frac{2}{\beta -4}}\left(\frac{r^2}{4 \theta }\right)}{8 \pi ^{3/2} (\beta -4) \theta ^{5/2}},\label{35}\\
F_h &=&\frac{1}{64 \pi ^{3/2} (\beta -4)^3 (\beta +2) \theta ^{7/2} r^5}\bigg(512 \pi ^{3/2} \alpha  (\beta -3) (\beta -2) C_{1} \theta ^{7/2} ((\beta -4) (-r))^{-\frac{4}{\beta -4}}\nonumber\\&+&(\beta -4)^2 (\beta +2) M r^6\bigg(3 \beta  r^2 E_{-3-\frac{2}{\beta -4}}\left(\frac{r^2}{4 \theta }\right)-4 (5 \beta +2) \theta  E_{-2-\frac{2}{\beta -4}}\left(\frac{r^2}{4 \theta }\right)\bigg)\bigg),\label{36}\\
F_e &=& \frac{\beta  \left(\frac{(\beta -4)^2 (\beta +2) M r^6 \left(28 \theta  E_{-2-\frac{2}{\beta -4}}\left(\frac{r^2}{4 \theta }\right)-3 r^2 E_{-3-\frac{2}{\beta -4}}\left(\frac{r^2}{4 \theta }\right)\right)}{\pi ^{3/2} \theta ^{7/2}}-256 \alpha  (\beta -3) C_{1} ((\beta -4) (-r))^{-\frac{4}{\beta -4}}\right)}{64 (\beta -4)^3 (\beta +1) r^5}.\label{37}
\end{eqnarray}

For the Lorentzian distribution the $F_a$, $F_h$ and $F_e$ forces are calculated as
\begin{eqnarray}
F_a &=&\frac{4}{\pi ^2 (\beta -4)^2 r^5}\bigg((\beta -4) r \left(\frac{\pi ^2 \alpha  (\beta -3) C_{2} ((\beta -4) (-r))^{-\frac{\beta }{\beta -4}}}{\beta +2}-\frac{(\beta -1) \sqrt{\theta } M r^3}{\left(\theta +r^2\right)^2}\right)\nonumber\\&+&2 (\beta -3) (\beta -1) \sqrt{\theta } M \left(-\frac{r^2}{\theta }\right)^{-\frac{2}{\beta -4}} \left(B_{-\frac{r^2}{\theta }}\left(1+\frac{2}{\beta -4},-1\right)-B_{-\frac{r^2}{\theta }}\left(1+\frac{2}{\beta -4},0\right)\right)\bigg),\label{38}\\
F_h &=&\frac{4}{\pi ^2 (\beta -4)^2 r^4}\bigg(-\frac{2 \pi ^2 \alpha  (\beta -3) (\beta -2) C_{2} ((\beta -4) (-r))^{-\frac{\beta }{\beta -4}}}{\beta +2}+\frac{1}{\left(\theta +r^2\right)^3}\bigg(\sqrt{\theta } M r^3 \left(2 (\beta -2) \right.\nonumber \\&+&\left.(\beta -1) \theta (\beta  (5 \beta -18)+4) r^2\right)\bigg)+\frac{4 (\beta -3) (\beta -1) M r}{\sqrt{\theta }}\bigg(-\, _2F_1\left(1,1+\frac{2}{\beta -4};2+\frac{2}{\beta -4};-\frac{r^2}{\theta }\right)\nonumber \\&\times&\, _2F_1\left(2,\frac{2}{\beta -4}+1;\frac{2}{\beta -4}+2;-\frac{r^2}{\theta }\right)\bigg)\bigg),\label{39}\\
F_e &=& \frac{2\beta}{\pi ^2 (\beta -4)^2 (\beta +1) r^5}\bigg(\pi ^2 \alpha  (\beta -3) C_{2} r ((\beta -4) (-r))^{-\frac{\beta }{\beta -4}}-\frac{(\beta +2) \sqrt{\theta } M r^4 \left((\beta -1) \theta +(4 \beta -13) r^2\right)}{\left(\theta +r^2\right)^3}\nonumber\\&+&\frac{2 (\beta -3) \left(\beta ^2+\beta -2\right) \sqrt{\theta } M \left(-\frac{r^2}{\theta }\right)^{-\frac{2}{\beta -4}} \left(B_{-\frac{r^2}{\theta }}\left(1+\frac{2}{\beta -4},-1\right)-B_{-\frac{r^2}{\theta }}\left(1+\frac{2}{\beta -4},0\right)\right)}{\beta -4}\bigg).\label{40}
\end{eqnarray}

\begin{figure}
\centering \epsfig{file=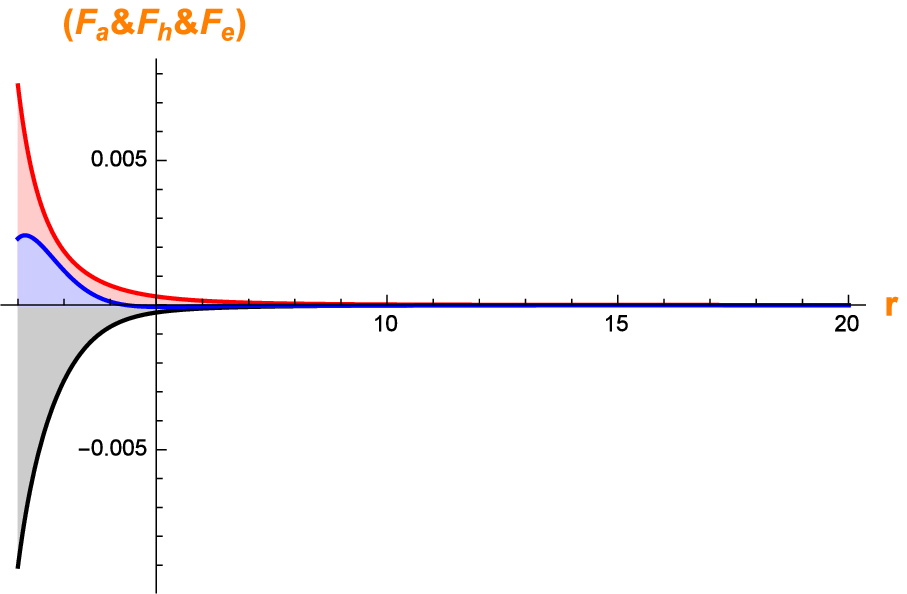, width=.32\linewidth,
height=2.02in}\epsfig{file=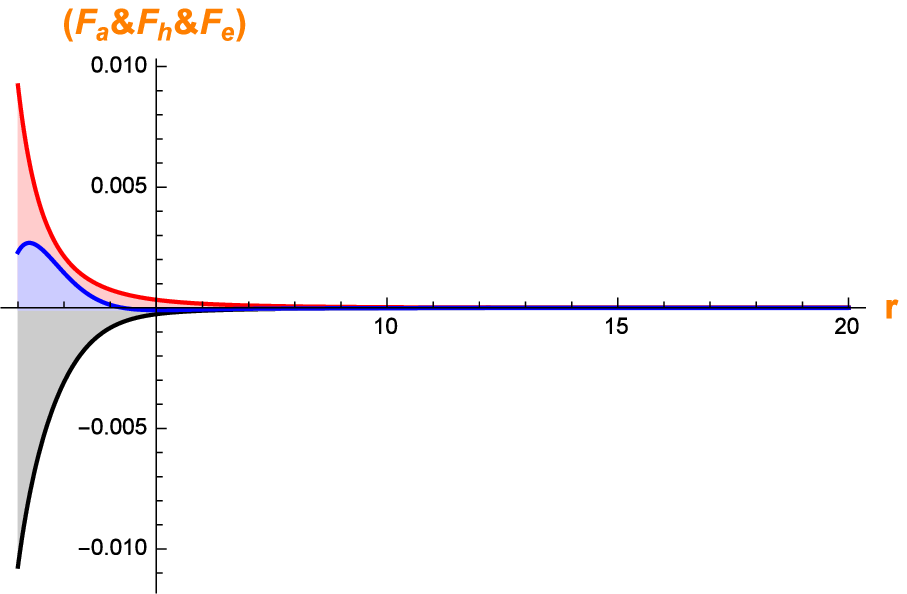, width=.32\linewidth,
height=2.02in}\epsfig{file=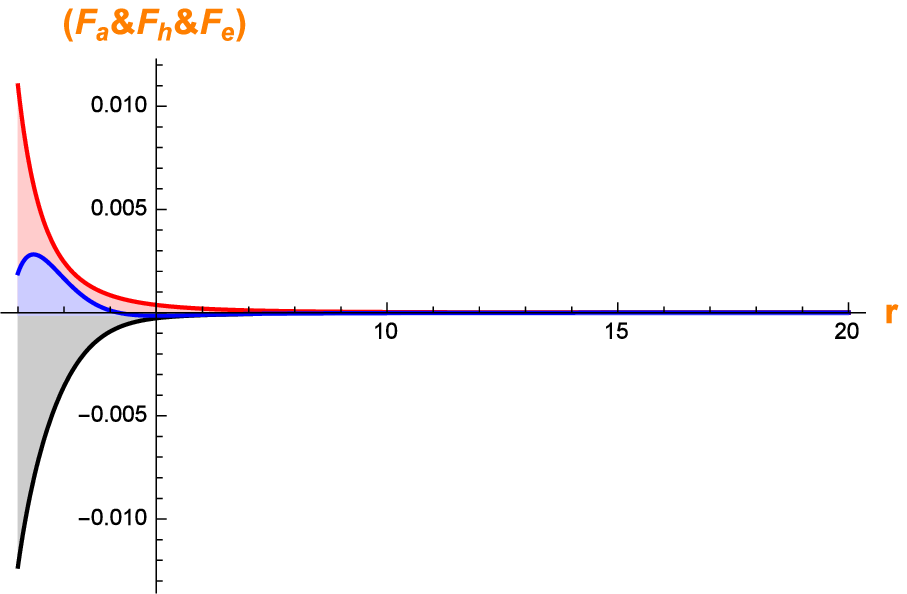, width=.32\linewidth,
height=2.02in}
\centering \epsfig{file=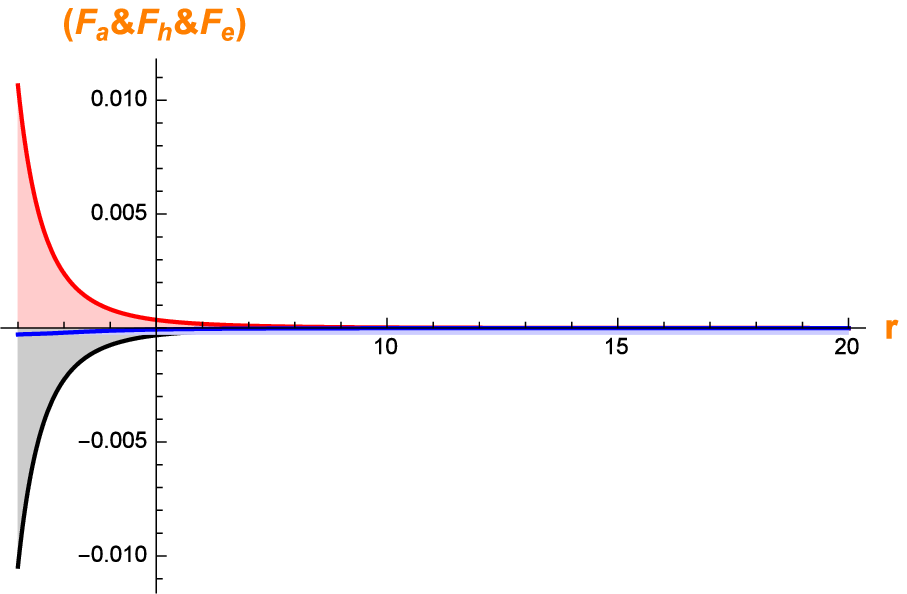, width=.32\linewidth,
height=2.02in}\epsfig{file=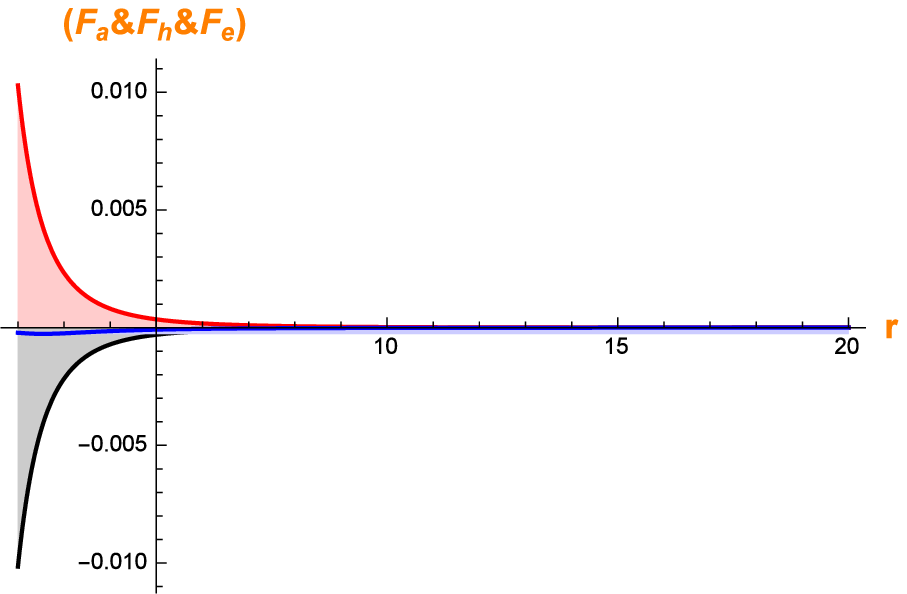, width=.32\linewidth,
height=2.02in}\epsfig{file=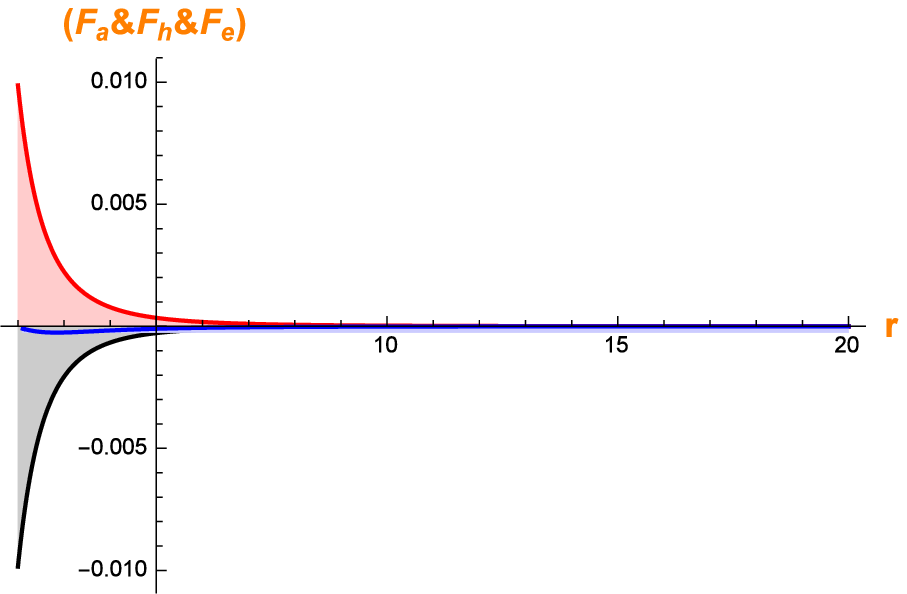, width=.32\linewidth,
height=2.02in} \caption{\label{fig10} Shows the behavior of TOV equation}
\end{figure}
Fig. (\ref{10}) provides the balanced behavior of the diversified forces $F_a$, $F_h$ and $F_e$. The first row of Fig. (\ref{10}) shows that these forces balance the effect of each other and leave the stable WH configuration under the particular values of different involved parameters for the Gaussian distribution. The second row of Fig. (\ref{10}) gives the stable WH configuration under similar conditions for the Lorentzian distribution. The balanced behavior of the three different forces shows that our obtained results are physically viable for the existence of WH geometries in both the cases. A detailed analysis of these forces is provided in Tabs. (\ref{V}-\ref{VI}).
\begin{center}
\begin{table}
\caption{\label{V}{Detailed summary of $F_a$, $F_h$ and $F_e$ for the Gaussian non-commutative distribution under $\alpha =0.5$, $C_{1}=0.2$, $\theta =0.9$, $ M=0.5$, and $\psi =3.036\times10^{-34}$}}
\begin{tabular}{|c|c|c|c|c|c|c|c|c|}
    \hline
\multicolumn{4}{|c|}{Gaussian Noncommutative Distribution}\\
    \hline
$Parameter/Expressions$    & $\beta=0.70$          &$\beta=0.90$  &$\beta=1.10$\\
\hline
$F_h$            & $F_h<0$ in $0.1\leq r\leq20$   &$F_h<0$ in $0.1\leq r\leq20$  &$F_h<0$ in $0.1\leq r\leq20$ \\

$F_e$            & $F_e>0$ in $0.1\leq r\leq20$   &$F_e>0$ in $0.1\leq r\leq20$ &$F_e>0$ in $0.1\leq r\leq20$\\

$F_a$            & $F_a>0$ in $0.1\leq r\leq20$   &$F_a>0$ in $0.1\leq r\leq20$  &$F_a>0$ in $0.1\leq r\leq20$ \\

$F_{a},\&F_{h},\& F_e$  &$F_{a},\&F_{h},\& F_e$(Balanced) &$F_{a},\&F_{h},\& F_e$(Balanced)  &$F_{a},\&F_{h},\& F_e$(Balanced)\\
\hline
\end{tabular}
\end{table}
\end{center}

\begin{center}
\begin{table}
\caption{\label{VI}{Detailed summary of $F_a$, $F_h$ and $F_e$ for Lorentzian non-commutative distribution under $\alpha =0.5$, $C_{1}=0.2$, $\theta =0.9$, $ M=0.5$, and $\psi =3.036\times10^{-34}$}}
\begin{tabular}{|c|c|c|c|c|c|c|c|c|}
    \hline
\multicolumn{4}{|c|}{Lorentzian Noncommutative Distribution}\\
    \hline
$Parameter/Expressions$    & $\beta=0.70$          &$\beta=0.90$  &$\beta=1.10$\\
\hline
$F_h$            & $F_h<0$ in $0.1\leq r\leq20$   &$F_h<0$ in $0.1\leq r\leq20$  &$F_h<0$ in $0.1\leq r\leq20$ \\

$F_e$            & $F_e<0$ in $0.1\leq r\leq20$   &$F_e<0$ in $0.1\leq r\leq20$ &$F_e<0$ in $0.1\leq r\leq20$\\

$F_a$            & $F_a>0$ in $0.1\leq r\leq20$   &$F_a>0$ in $0.1\leq r\leq20$  &$F_a>0$ in $0.1\leq r\leq20$ \\

$F_{a},\&F_{h},\& F_e$  &$F_{a},\&F_{h},\& F_e$(Balanced) &$F_{a},\&F_{h},\& F_e$(Balanced)  &$F_{a},\&F_{h},\& F_e$(Balanced)\\
\hline
\end{tabular}
\end{table}
\end{center}

\section{Embedding diagram of wormhole with Gaussian and Lorentzian distributions}

To symbolize the noncommutative WH structure, we need to discuss the embedding figure and extract the specifically required conditions. To take specific spherical symmetric space-time with an equatorial slice, we use $\theta=2\pi$ and $t = const.$ in Eq. (\ref{3}), which then becomes
\begin{equation} \label{41}
ds^{2}= \left(1-\frac{\epsilon _s(r)}{r}\right)^{-1}dr^{2}+r^{2}d\phi^{2},
\end{equation}
The Eq. (\ref{3}) can be embedded into 3-D Euclidean space-time with cylindrical symmetry, which is expressed as
\begin{equation} \label{42}
ds^{2}_{\Xi}= dh^{2}+dr^{2}+r^{2}d\phi^{2},
\end{equation}
The above Eq. (\ref{42}) can be rewritten as
\begin{equation} \label{43}
ds^{2}_{\Xi}= \bigg(1+\bigg(\frac{dh}{dr}\bigg)^{2}\bigg)dr^{2}+r^{2}d\phi^{2},
\end{equation}
By matching Eqs. (\ref{41}-\ref{43}), we get the following relation
\begin{equation} \label{43}
\frac{dh}{dr}=\pm \left(\frac{r}{\epsilon _s(r)}-1\right)^{-1/2},
\end{equation}

\begin{figure}
\centering \epsfig{file=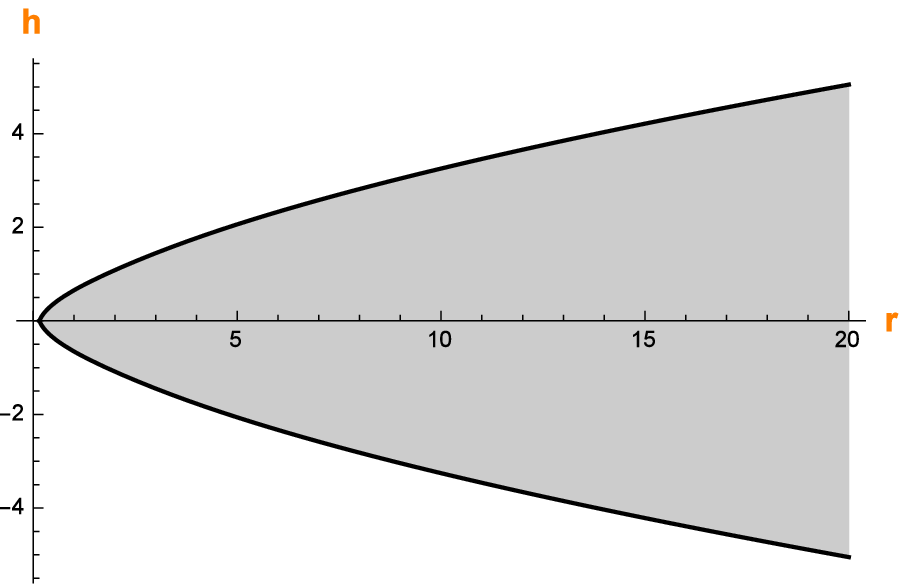, width=.32\linewidth,
height=2.02in}\epsfig{file=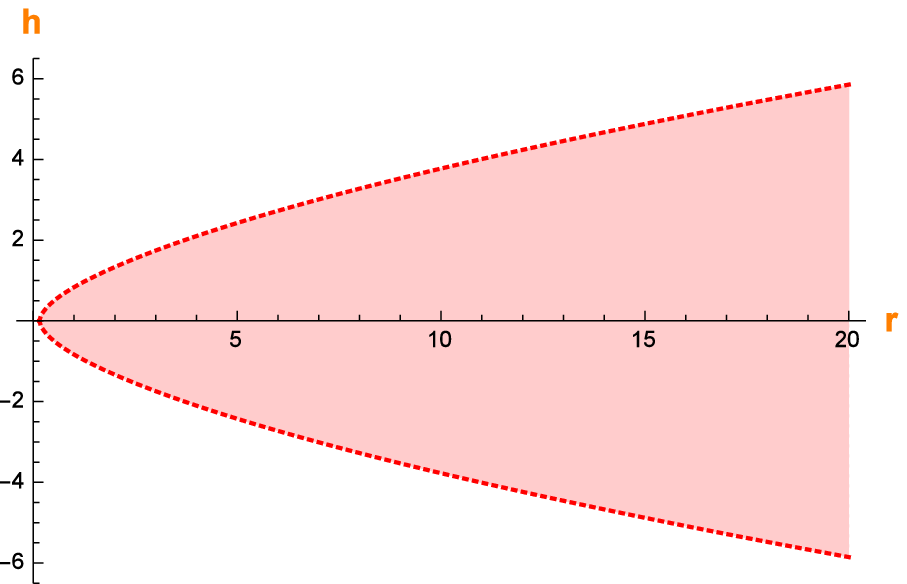, width=.32\linewidth,
height=2.02in}\epsfig{file=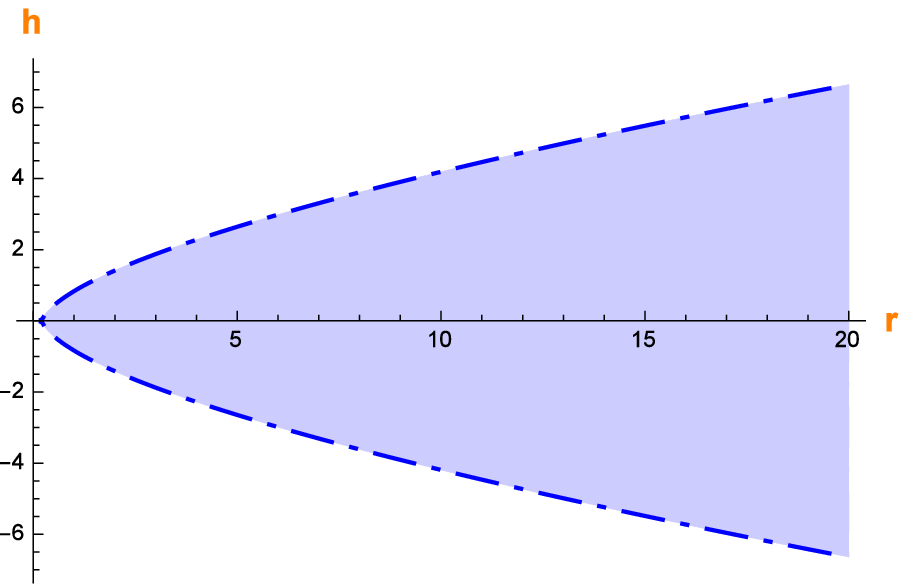, width=.32\linewidth,
height=2.02in}
\centering \epsfig{file=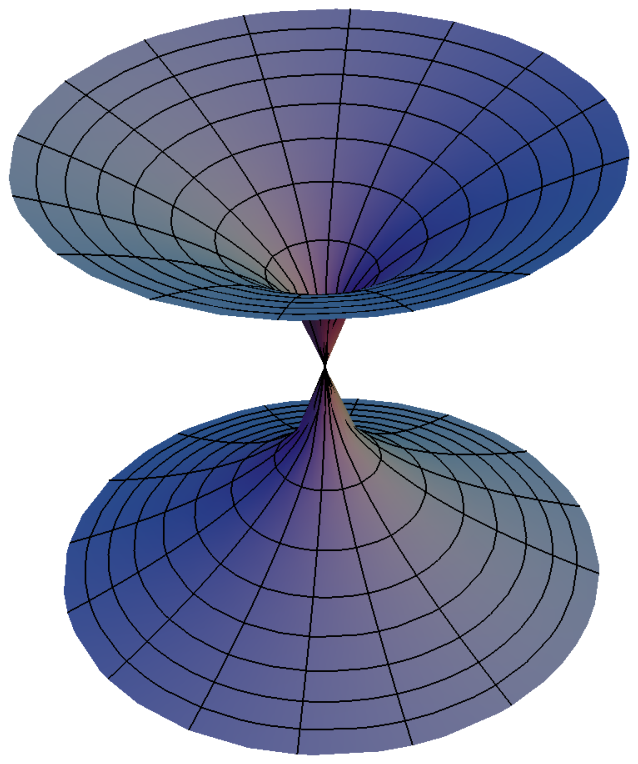, width=.32\linewidth,
height=2.02in}\epsfig{file=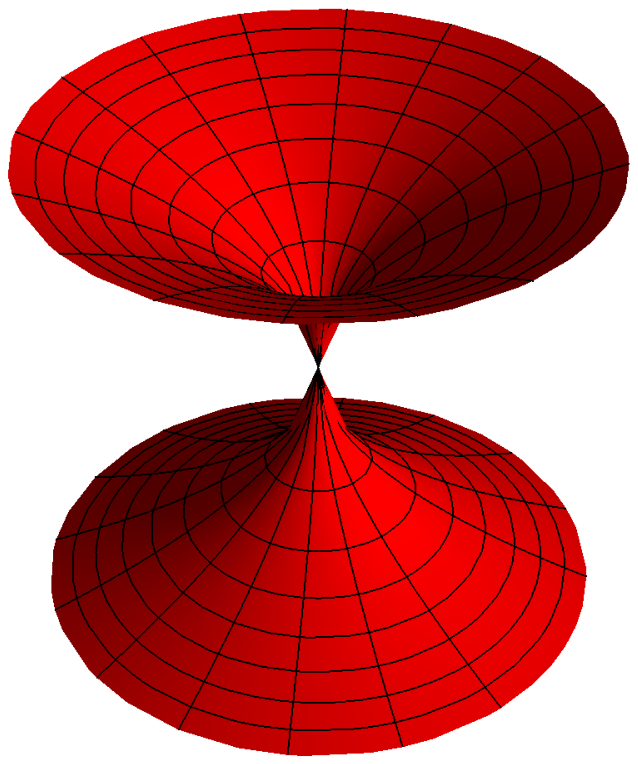, width=.32\linewidth,
height=2.02in}\epsfig{file=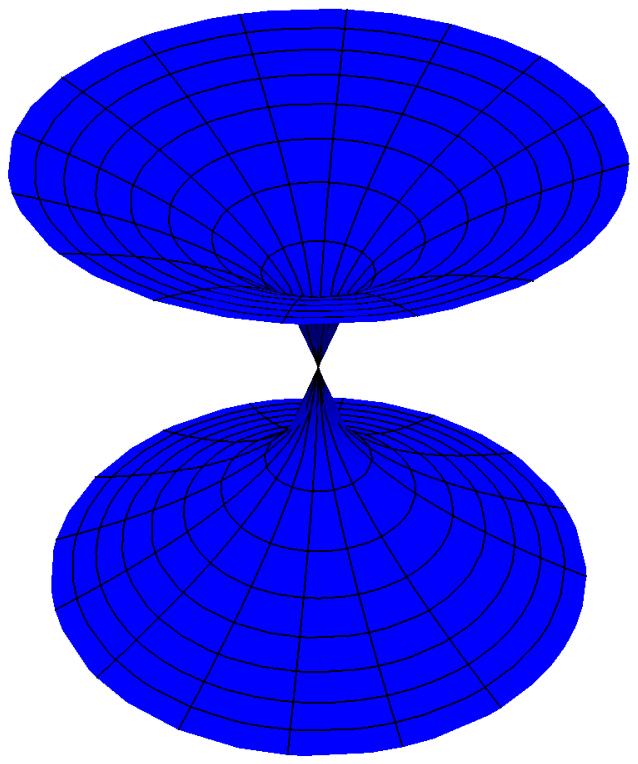, width=.32\linewidth,
height=2.02in} \caption{\label{fig11} Shows the behavior of embedding diagram for lower and upper Universe.}
\end{figure}

\begin{figure}
\centering \epsfig{file=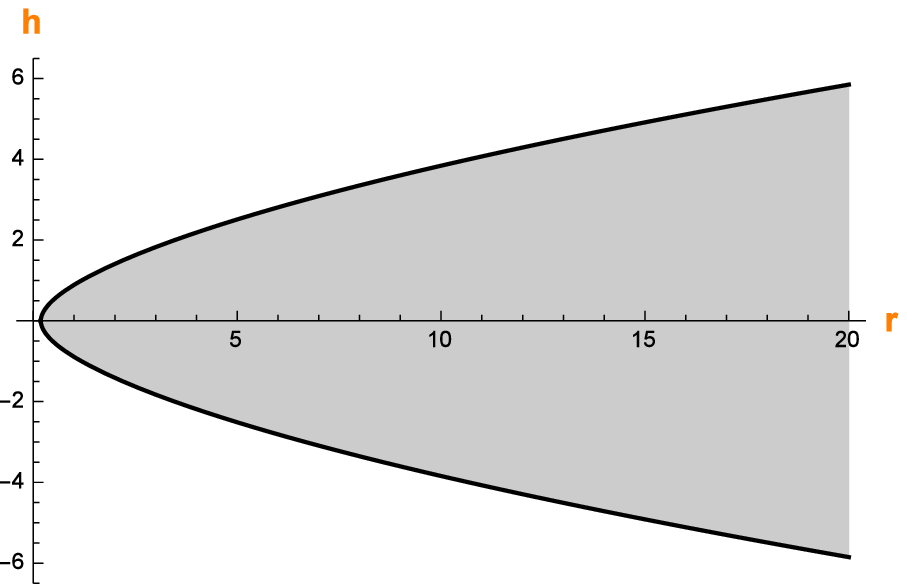, width=.32\linewidth,
height=2.02in}\epsfig{file=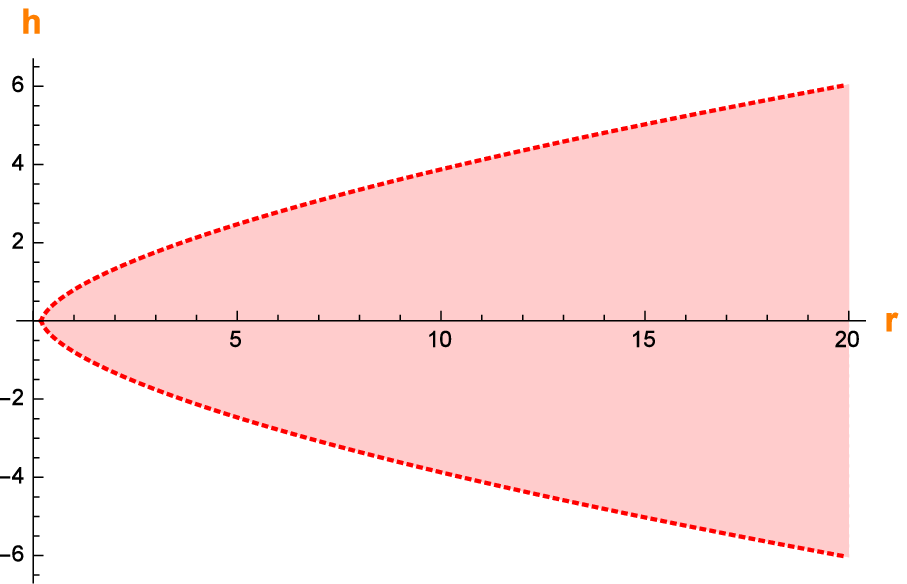, width=.32\linewidth,
height=2.02in}\epsfig{file=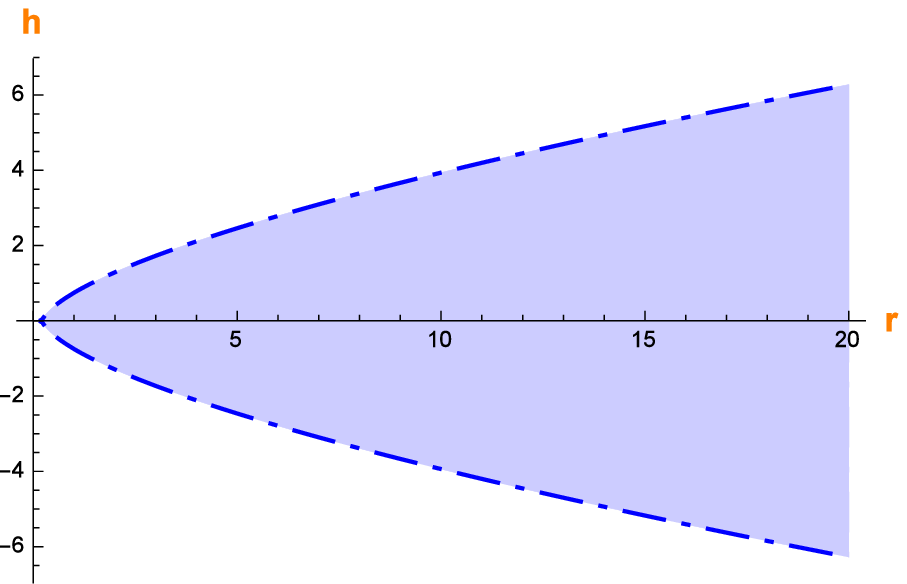, width=.32\linewidth,
height=2.02in}
\centering \epsfig{file=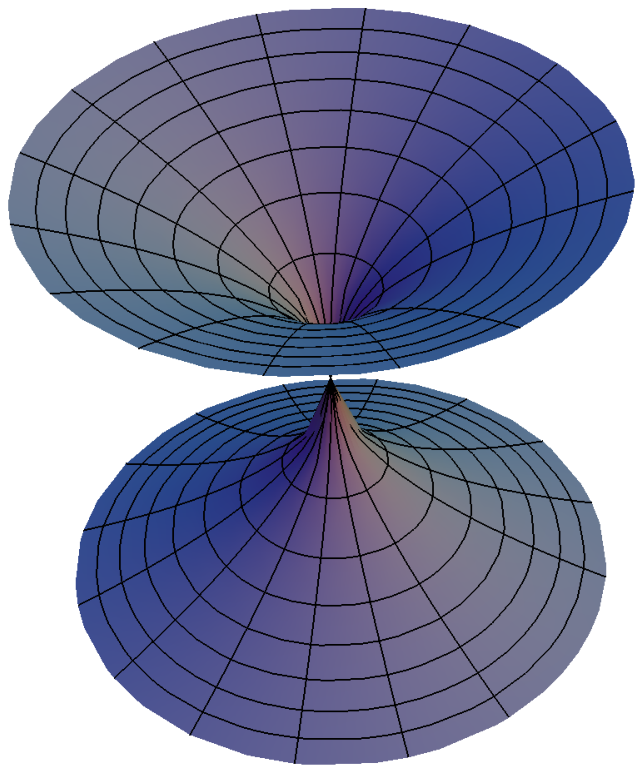, width=.32\linewidth,
height=2.02in}\epsfig{file=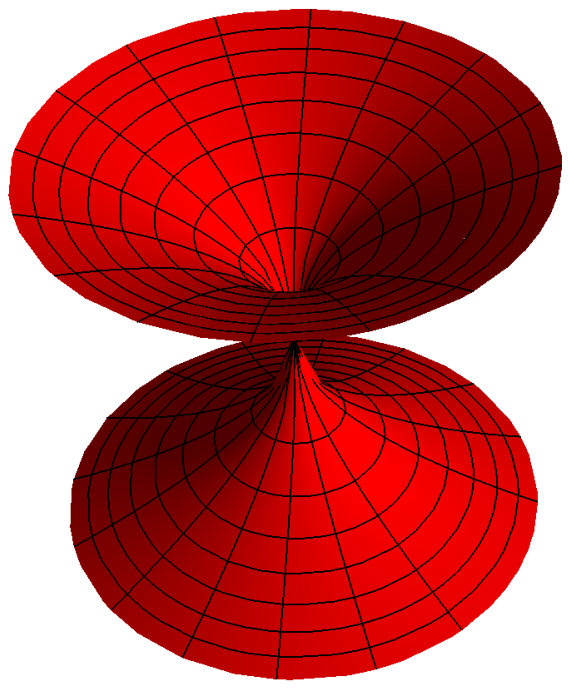, width=.32\linewidth,
height=2.02in}\epsfig{file=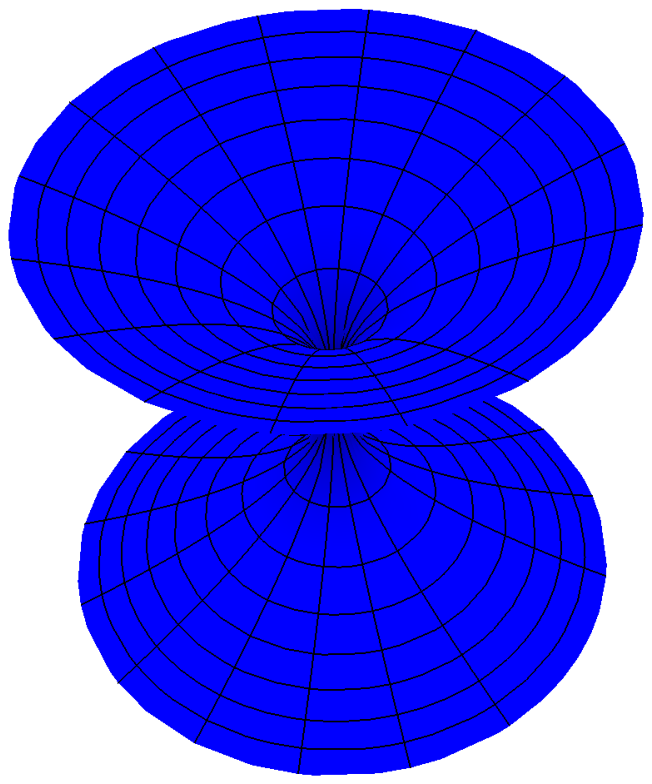, width=.32\linewidth,
height=2.02in} \caption{\label{fig12} Shows the behavior of embedding diagram for lower and upper Universe. }
\end{figure}
From Eq. (\ref{43}) the expression $\frac{dh}{dr}$ approaches to zero as $\frac{\epsilon _s}{r}\rightarrow 0$ with $r\rightarrow\infty$ shows that at the throat $r_{0}$ the embedded surface should be vertical. The embedded surface diagram for $h(r)>0$ (upper universe) and $h(r)<0$ (lower universe) for the Gaussian distribution can be seen from Fig. (\ref{11}). For the Lorentzian distribution, the embedded surface diagram for $h(r)>0$ (upper universe) and $h(r)<0$ (lower universe) is provided in Fig. (\ref{12}). Away from the $r_{0}$, the space is asymptotically flat because $\frac{dh}{dr}\rightarrow 0$ with $r\rightarrow\infty$, it can be confirmed from the embedded surface diagrams for both the distributions.

\section{Critical analysis on anisotropic pressure under Gaussian and Lorentzian distributions}
In this section, we shall discuss some aspects of anisotropic pressure. The Gaussian and Lorentzian energy densities by Eq. (\ref{12}) should remain positive under the positive values of involved parameters, i.e., $\theta =0.9$ and $ M=0.5$. In this study, the anisotropy, i.e., $\triangle=p_{t}-p_{r}$ is seen positive. The positive behavior of $\triangle$ shows that $p_{t}>p_{r}$, which can be verified from the Fig. (\ref{13}). It is also found from Fig. (\ref{13}) that $\mid p_{t}\mid<\mid p_{r}\mid$. Further, the positive behavior of $\triangle$ guarantees the presence of exotic matter. Comprehensive consolidated readings on pressure components and energy density are provided in Tab. (\ref{VII}).

\begin{center}
\begin{table}
\caption{\label{VII}{Detailed summary of $\rho$, $p_{r}$ and $p_{t}$ for the Gaussian non-commutative distribution under $\alpha =0.5$, $C_{1}=0.2$, $\theta =0.9$, $ M=0.5$, and $\psi =3.036\times10^{-34}$}}
\begin{tabular}{|c|c|c|c|c|c|c|c|c|}
    \hline
\multicolumn{4}{|c|}{Gaussian Non-commutative Distribution}\\
    \hline
$Parameter/Expressions$    & $\beta=0.70$          &$\beta=0.90$  &$\beta=1.10$\\
\hline
$\rho$            & $\rho<0$ in $0.1\leq r\leq20$   &$\rho<0$ in $0.1\leq r\leq20$  &$\rho<0$ in $0.1\leq r\leq20$ \\

$p_{r}$            & $p_{r}<0$ in $0.1\leq r\leq20$   &$p_{r}<0$ in $0.1\leq r\leq20$ &$p_{r}<0$ in $0.1\leq r\leq20$\\

$p_{t}$            & $p_{t}>0$ in $0.1\leq r\leq20$   &$p_{t}>0$ in $0.1\leq r\leq20$  &$p_{t}>0$ in $0.1\leq r\leq20$ \\

$\triangle$       & $\triangle>0$ in $0.1\leq r\leq20$   &$\triangle>0$ in $0.1\leq r\leq20$  &$\triangle>0$ in $0.1\leq r\leq20$ \\

$\frac{\rho}{p_{r}}$  & $\frac{\rho}{p_{r}}<0$ in $0.1\leq r\leq20$   &$\frac{\rho}{p_{r}}<0$ in $0.1\leq r\leq20$ &$\frac{\rho}{p_{r}}<0$ in $0.1\leq r\leq20$\\

$\frac{\rho}{p_{t}}$  & $\frac{\rho}{p_{t}}>0$ in $0.1\leq r\leq20$   &$\frac{\rho}{p_{t}}>0$ in $0.1\leq r\leq20$  &$\frac{\rho}{p_{t}}>0$ in $0.1\leq r\leq20$ \\
\hline
\end{tabular}
\end{table}
\end{center}

\begin{figure}
\centering \epsfig{file=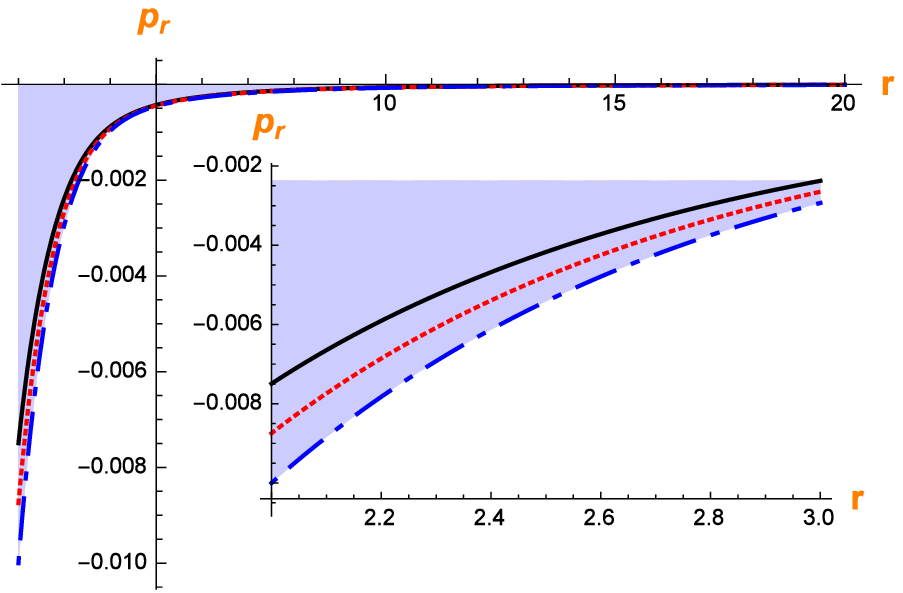, width=.32\linewidth,
height=2.02in}\epsfig{file=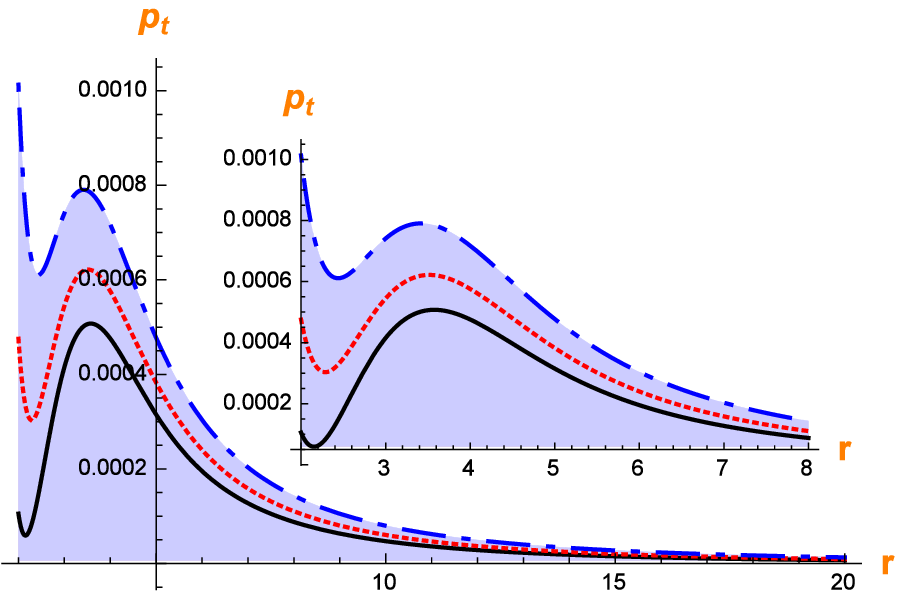, width=.32\linewidth,
height=2.02in}\epsfig{file=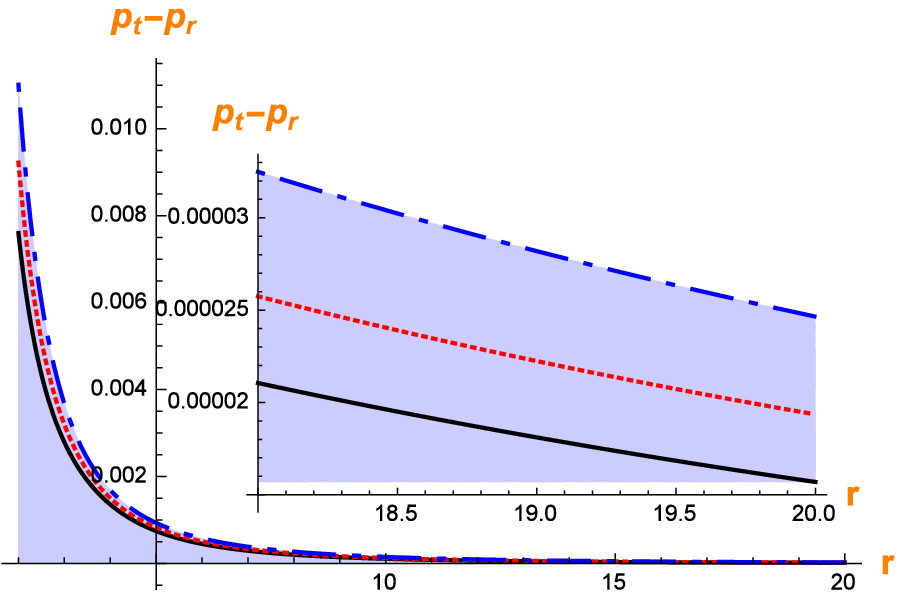, width=.32\linewidth,
height=2.02in}
\centering \epsfig{file=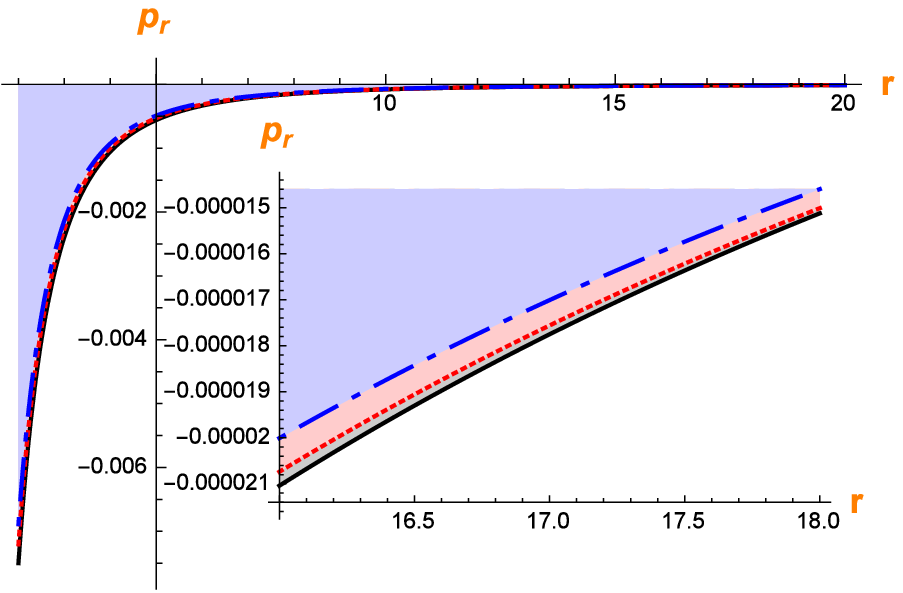, width=.32\linewidth,
height=2.02in}\epsfig{file=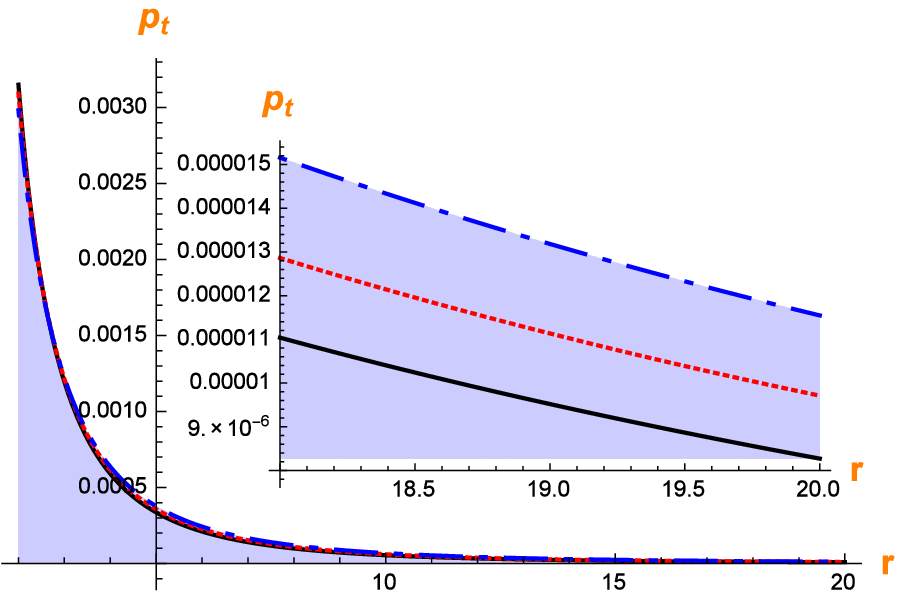, width=.32\linewidth,
height=2.02in}\epsfig{file=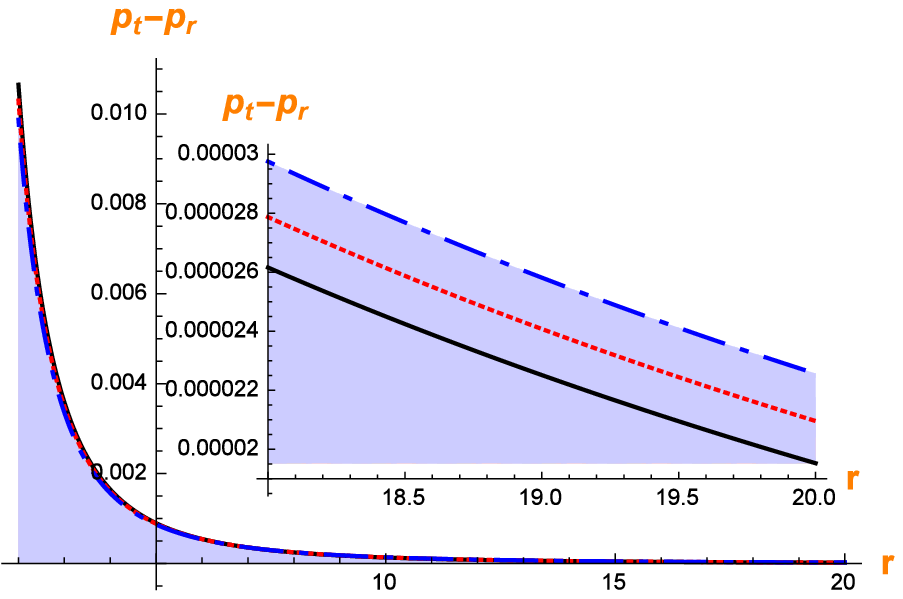, width=.32\linewidth,
height=2.02in} \caption{\label{fig13} Shows the behavior of anisotropy and pressure components.}
\end{figure}

\section{Conclusion}

In GR, the existence of WH geometries possessing exotic matter has always fascinated the researchers
as its presence leads to the violation of  NEC which then guarantees the existence of WH geometries.
 However, as for the modified theories are concerned,
finding the WH solutions  becomes even more captivating
topic due to the inclusion of effective energy-momentum
tensor that causes the violation of NEC regardless of the presence of any such separate exotic
matter. In this work, we have explored the
spherically symmetric WH geometries involving
the Gaussian and the Lorentzian non-commutative sources
 in the modified  $f(T,\tau)$ gravity. To achieve our goal, we have incorporated a linear model, i.e., $ f(\tau,T)=\alpha\tau(r)+\beta T+\phi $, where  $\alpha,\;\beta$ and $\phi$ are constants, and $\tau(r)=\frac{2 e^{-b (r)}}{r^{2}}$. Using the energy-momentum tensor and the other crucial ingredients, we have worked out the exclusive expressions for the energy density, radial, and tangential components of pressure.

We have worked out the shape function $\epsilon _s(r)$ for both the Gaussian and the Lorentzian distributions separately with three different values of the matter coupling parameter, i.e., $\beta=0.70,\;0.90,\;\&\;1.10$ and have analyzed them graphically. It can be verified from Fig. (\ref{1}) that $\epsilon _s(r)$ smoothly increases and remains positive for both the cases. Such behavior of the $\epsilon _s(r)$ displays that our calculated shape functions in both the Gaussian and the Lorentzian distributions support the existence of WH geometries in our work. We have also examined the tangents of the shape function $\epsilon _s(r)$ for both the sources with the same values of the parameter $\beta$ and have found that $\frac{d\epsilon _s}{dr}<1$ which is evident the from Fig. (\ref{2}).  The critical constraint of $\frac{d\epsilon _s}{dr}<1$  describes the flaring out the condition of WH geometries which is justified in both of our cases to favor the existence of the WH geometries as well.\\

We have also discussed for both the Gaussian and the Lorentzian distributions an important ratio $\frac{\epsilon _s}{r}\rightarrow 0$ as $r\rightarrow\infty$  which gives the flatness property to be fulfilled for the existence of valid WH geometries. The plots in Fig. (\ref{3}) tell us the story of the fulfillment of this critical condition for the parametric values of $\beta=0.70,\;0.90,\;\&\;1.10$ The flatness conduct demonstrates the overwhelming character of non-commutative geometry in the WH study. We have worked out the WH throats locations through $\epsilon _{s}-r$ for both the Gaussian and the Lorentzian distributions. From the plots of Fig. (\ref{4}), the throat positions can be sited against the diversified parametric values of $\beta$,i.e., $\beta=0.70,\;0.90,\;\&\;1.10$. In the Gaussian framework, the WH throats are premeditated as $r_{0}=0.120$ for $\beta=0.70$, $r_{0}=0.145$ for $\beta=0.90$, and $r_{0}=0.165$ for $\beta=1.10$, which can be witnessed at the first row of the plots of Fig. (\ref{4}). For the Lorentzian distribution, the WH throats locations are found as $r_{0}=0.180$ for $\beta=0.70$, $r_{0}=0.165$ for $\beta=0.90$, and $r_{0}=0.145$ for $\beta=1.10$, these throats can be confirmed from the second row of Fig. (\ref{4}). These diverse values of WH throat reflect the critical impact of parameter $\beta$ in the current situation. The distinct values of the parameter $\beta$ deliver the diversified WH throat locations. Tables. (\ref{I}-\ref{II}) are confined with the thorough aspects for both the cases under investigations. \\

The role of the Energy bounds has always remained very critical to explore the WH geometries. For this purpose, we have worked out the expressions $\rho+p_{r}$, $\rho-p_{r}$, $\rho+p_{t}$, $\rho-p_{t}$, and $\rho+p_{r}+2p_{t}$, to constitute the energy constraints. We have analyzed the behavior of ECs as reflected in the plots of Figs. (\ref{5}-\ref{9}) for both, the Gaussian and the Lorentzian distributions. Fig. (\ref{5}), unveils the violation of $NEC$ i.e., $\rho+p_{r}<0$ within the radial constraint of $1\leq r\leq20$ for $\beta=0.70,\;0.90,\;\&\;1.10$, hence favoring both of the non-commutative geometries. The $(NEC)$ violation is the key to the presence of exotic matter in both the cases and is essential requirement for the existence of WH solutions. We have noted the positive profile of $\rho-p_{r}$, $\rho+p_{t}$, and $\rho+p_{r}+2p_{t}$ as shown in Figs. \ref{6},\ref{7}, and \ref{9} under the same parametric conditions within $1\leq r\leq20$. Moreover, we have also witnessed the negative behavior of $\rho-p_{t}$ as reflected in Fig. (\ref{8}) for both of the distributions. The results concerning ECs are provided in Tables. (\ref{III}-\ref{IV}).

We have also investigated the stability of our emerging solutions by incorporating the Tolman-Oppenheimer-Volkov (TOV) equation. For this purpose, we have worked out the constituent diverse forces  $F_a$, $F_h$, and $F_e$, the total effect of which remained almost zero to confirm the stability of WH configuration as reflected in the plots of Fig. (\ref{10}) for both the cases under the same conditions. A detailed analysis of the three different forces is provided in Tables. (\ref{V}-\ref{VI}).

To highlight the nature of the non-commutative WH structure, we have demonstrated specifically the embedding figures. The embedded surface diagram for $h(r)>0$ (upper universe) and $h(r)<0$ (lower universe) for the Gaussian and the Lorentzian distributions is reflected in Fig. (\ref{11}), and Fig. (\ref{12}), respectively. We note that beyond the positioning of $r_{0}$, the space is asymptotically flat due to $\frac{dh}{dr}\rightarrow0$ as $r\rightarrow\infty$.
Finally, we have also discussed some characteristics of anisotropic pressure. The Gaussian and the Lorentzian energy densities explained by Eq. (\ref{12}) have remained positive in our study for $\theta =0.9$ and $ M=0.5$. Moreover, the anisotropy, $\triangle=p_{t}-p_{r}$ has been noted as positive such that $p_{t}>p_{r}$, as shown in Fig. (\ref{13}). We have also noted that $\mid p_{t}\mid<\mid p_{r}\mid$. Further, the positive behavior of $\triangle$  warranties the presence of exotic matter. A detailed summary of pressure components and energy density can be seen in Table. (\ref{VII}).

\end{document}